%% file: ms.tex
\documentclass[12pt,preprint]{aastex}

\shorttitle{Testing the standard fireball model}
\shortauthors{Willingale et al.}

\begin{document}


\title{Testing the standard fireball model of GRBs using
late X-ray afterglows measured by {\em Swift}}


\author{R. Willingale\altaffilmark{1},
P.T. O'Brien\altaffilmark{1},
J.P. Osborne\altaffilmark{1},
O. Godet\altaffilmark{1},
K.L. Page\altaffilmark{1},
M.R. Goad\altaffilmark{1},
D.N. Burrows\altaffilmark{2},
B. Zhang\altaffilmark{4},
E. Rol\altaffilmark{1},
N. Gehrels\altaffilmark{3},
G. Chincarini\altaffilmark{5}
}


\altaffiltext{1}{Department of Physics and Astronomy, University of Leicester,
LE1 7RH, UK}
\altaffiltext{2}{Department of Astronomy and Astrophysics, Pennsylvania State
University, University Park, PA 16802, USA}
\altaffiltext{3}{NASA Goddard Space Flight Center, Greenbelt, Maryland, 20771, USA}
\altaffiltext{4}{Department of Physics, University of Nevada, Las Vegas, NV 89154,
USA}
\altaffiltext{5}{INAF, Osservatorio Astronomico di Brera, Via E. Bianchi 46,
I-23807 Merate (LC), Italy}


\begin{abstract}
We show that all X-ray decay curves of GRBs measured by {\em Swift} can
be fitted using one or two components both of which have exactly the same
functional form comprised of an early falling exponential phase followed
by a power law decay. The 1st component contains the prompt $\gamma$-ray
emission and the initial X-ray decay. The 2nd component appears
later, has a much longer duration and is present for $\approx80\%$ of GRBs. 
It most likely arises from the external shock which eventually
develops into the X-ray afterglow. In the remaining $\approx20\%$ of 
GRBs the initial X-ray decay of the 1st component fades more slowly than
the 2nd and dominates at late times to form an afterglow
but it is not clear what the origin of this emission is.

The temporal decay parameters and
$\gamma$/X-ray spectral indices derived for 107 GRBs 
are compared to the expectations of the standard fireball model
including a search for possible ``jet breaks''.
For $\sim50\%$ of GRBs the observed afterglow
is in accord with the model but for the rest the temporal and
spectral indices do not conform to the expected closure relations and 
are suggestive of continued, late, energy injection. We identify a few possible
jet breaks but there are many examples where such breaks are predicted
but are absent.

The time, $T_{a}$, at which the exponential phase of the
2nd component changes to a final
powerlaw decay afterglow is correlated with the peak of
the $\gamma$-ray spectrum, $E_{peak}$. This is analogous to
the Ghirlanda relation, indicating that
this time is in some way related to optically observed break times
measured for pre-{\em Swift} bursts. Many optical breaks have
previously been identified as jet breaks but the differences seen
between X-ray and optical afterglows suggest that this is not
the explanation.
\end{abstract}





\section{Introduction}

The standard fireball shock model of GRBs (M\'{e}sz\'{a}ros 2002 and references
therein) predicts that a broadband continuum afterglow spectrum is expected
to arise from  an external shock 
when the relativistically expanding fireball is decelerated by
the surrounding low density medium. As relativistic electrons,
accelerated in the shock to form a power law energy spectrum, spiral
in the co-moving magnetic field we should see a characteristic fading
synchroton radiation spectrum stretching from radio frequencies through the
IR, optical and UV bands in to an X-ray and gamma ray high energy tail.
The detailed form of the expected afterglow spectrum and its evolution
are described by Sari, Piran and Narayan (1998) and Wijers and Galama (1999).

X-ray afterglows of GRBs were first detected by the {\em Beppo-SAX} satellite 
(1996-2002) and the detection of GRB970228 (Costa et al. 1997) and
other X-ray afterglows provided positions of sufficient accuracy to
enable follow-up ground-based optical observations. Faint optical
afterglows were discovered and it was soon established that GRBs
occurred at comological distances. The first redshift, z=0.835,
was measured for GRB970508 (Metzger et al. 1997). A connection between
GRBs and supernovae was revealed by observations of GRB980425/SN1998bw
(Galama et al. 1998, Kulkarni et al. 1998) although the supernovae
associated with GRBs showed very high expansion velocities (tens of
thousands of kilometers per second) and were given a new classification of
hypernovae. The XMM-Newton observatory also detected X-ray afterglows.
In particular GRB030329 confirmed the hypernova connection
(Tiengo et al. 2003, Stanek et al. 2003, Hjorth et al, 2003)
and multiwavelength observations and analysis of this bright afterglow and 
similar events (Harrison et al. 2001, Panaitescu \& Kumar 2001, 2002, 2003,
Willingale et al. 2004) established that afterglows were broadly consistent
with the expected synchrotron spectrum and temporal evolution.

If the relativistic outflow is collimated in the form of a jet then we expect
to see an achromatic break in the decay at time $t_{j}$ days after the burst
when the edge of the jet becomes visible (Rhoads 1997, 1999).
Many optical observations of GRB afterglow decays 
exhibit a break a few days after the initial burst, which is identified with a
jet break,
consistent with a collimation angle, $\theta_{j}$ $\sim 3-40$
degrees (Frail et al. 2001, Bloom et al. 2003).
Assuming the fireball emits a fraction
$\eta_{\gamma}$ of its kinetic energy in the prompt $\gamma$-ray emission
and the circumburst medium has constant number density $n$ the 
collimation angle is given by
\begin{equation}
\theta_{j}=0.161\left(\frac{t_{j}}{1+z}\right)^{3/8}
\left(\frac{n\eta_{\gamma}}{E_{iso}}\right)^{1/8}
\label{eq1}
\end{equation}
where z is the redshift, and $E_{iso}$ is the total energy in $\gamma$-rays
in units of $10^{52}$ ergs calculated assuming the emission is isotropic
(Sari et al. 1999). The collimation-corrected energy is then
$E_{\gamma}=E_{iso}(1-\cos\theta_{j})$ and this shows a tight correlation
with the peak energy of the spectrum in the source-frame,
$E_{peak}^{src}\propto E_{\gamma}^{0.7}$ (the Ghirlanda relation:
Ghirlanda et al. 2004). Jet breaks seen in the optical should
also be observed, simultaneously, in the X-ray band.

Prior to the launch of {\em Swift} (Gehrels et al. 2004, Burrows et al. 2005)
both X-ray and optical follow-up observations
of GRBs and their afterglows were limited to late times greater than several 
hours and often a day or more after the GRB trigger. Since launch, {\em Swift}
has detected an average of 2 GRBs per week and we now have a sample
of over 100 GRBs for which we have quasi-continuous coverage in the X-ray band
in the range $\sim100$ to $\sim10^{6}$ seconds after the initial trigger.
The aim of this paper is to compare the observed X-ray afterglows with
the expectations of the standard model. One possible approach is to correlate
the behaviour seen in the X-ray band with simultaneous optical measurements.
Panaitescu et al. (2006) present an analysis for 6 GRBs detected by {\em Swift}
noting that, contrary to expectation, temporal breaks in the X-ray band
were not seen simultaneously in the optical.
Another approach is to use the Ghirlanda relation
to predict the time of expected jet breaks for {\em Swift} GRBs for which we
have redshifts and then look to see if such breaks were observed.
Sato et al. (2006) applied this to 3 bursts without success and concluded
these bursts indicate a large scatter in the Ghirlanda relation.
A third possible approach, presented here, is
to make a systematic statistical study of the structure and evolution
of a large sample of X-ray decay curves (including the method employed
by Sato).

\section{The functional form of X-ray decays seen by {\em Swift}}

\begin{figure}[!htp]
\begin{center}
\includegraphics[height=15cm,angle=-90]{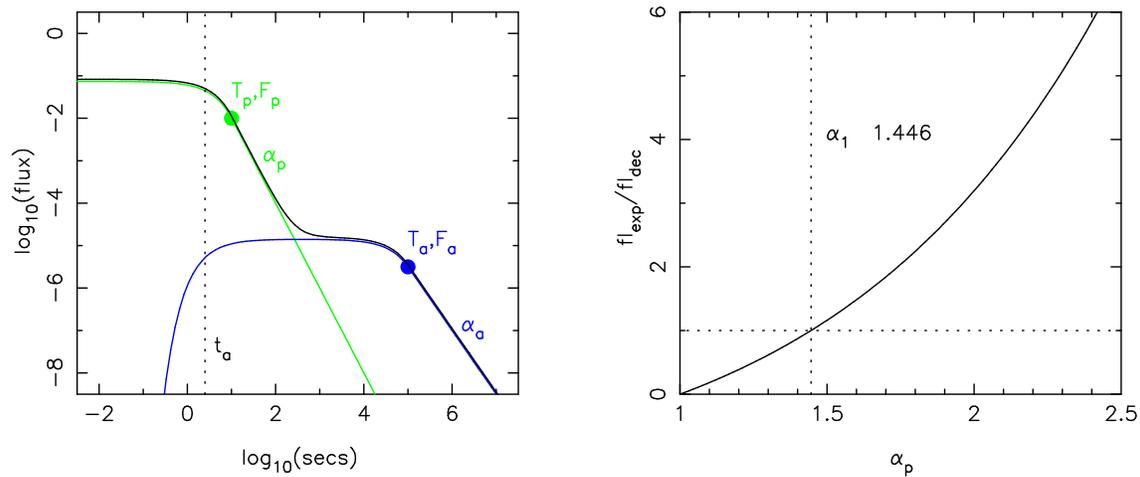}
\end{center}
\caption{Left: The functional form of the decay and the fitted parameters.
The prompt component (green) has no rise because time zero is set at the
peak. The afterglow component (blue) rises at time $t_{a}$ as shown.
Right: the ratio of the fluence in the exponential portion of the light
curve to that in the power law decay as a function of $\alpha_{p}$.
See Equations \ref{eq3}, \ref{eq4} and \ref{eq6}.}
\label{fig1}
\end{figure}

An analysis of a sample of 40 X-ray decays observed by {\em Swift} by
O'Brien et al. (2006), demonstrated that they all followed a similar pattern
comprising an exponential decay in the prompt phase which relaxes to
a power law decay at a time $T_{p}$. In most cases this initial power law
decay flattens into a plateau or shallow decay
which then gradually steepens and 
establishes a final afterglow power law decay at time $T_{a}$. Fig.
\ref{fig1} shows a schematic of the decay profile and the disposition of
$T_{p}$ and $T_{a}$.
Such behaviour is consistent with the presence of two emission components
that overlap in time;
a short duration prompt emission followed by an initial power law decay and
designated by the subscript ``p'' and
a much longer duration low luminousity afterglow component which starts
as a slowly decaying plateau and ends with a steeper powerlaw, designated by
the subscript ``a''.
The analysis reported by O'Brien et al. (2006) concentrated on the properties
of the prompt ``p'' component and produced an estimate of $T_{p}$ using
a scaled version of each X-ray light curve. In this paper we turn our
attention to the later development of X-ray light curves and employ
a function fitting procedure to estimate the parameters
associated with both the prompt and afterglow components.

We have found that both components are well fitted by the same functional form:
\[
f_{c}(t)=F_{c}
\exp\left(\alpha_{c}-\frac{t\alpha_{c}}{T_{c}}\right)
\exp\left(\frac{-t_{c}}{t}\right)
,\:\:t<T_{c}
\]

\begin{equation}
f_{c}(t)=F_{c} \left( \frac{t}{T_{c}}\right)^{-\alpha_{c}}
\exp\left(\frac{-t_{c}}{t}\right)
,\:\:t\ge T_{c}
\label{eq2}
\end{equation}
The transition from the exponential to the power law occurs at the point
$(T_{c},F_{c})$ where the two functional sections have the same value
and gradient. The parameter $\alpha_{c}$ determines both the time
constant of the exponential decay, $T_{c}/\alpha_{c}$, and the
temporal decay index of the power law. The time $t_{c}$ marks the 
the initial rise and the maximum flux
occurs at $t=\sqrt{t_{c}T_{c}/\alpha_{c}}$.

Having established this generic behaviour we have fitted the X-ray decay curves
of all 107 GRBs detected by both the BAT and XRT on {\em Swift} up to
August 1st 2006
using two components of the form $f(t)=f_{p}(t)+f_{a}(t)$. Parameters
with suffix $p$ ($T_{p}$,\ldots) refer to the prompt component
and those with suffix $a$ ($T_{a}$,\ldots) the afterglow component.
Fig. \ref{fig1} illustrates the functional form of the two components.

The X-ray light curves were formed from the combination of BAT and XRT
data as described by O'Brien et al. (2006).
The conventional prompt emission, seen predominantly by the BAT,
occurs for $t<T_{p}$ and the
plateau/shallow decay phase, seen by the XRT, at $t<T_{a}$.
The fits were produced in two stages. The first stage used 
the BAT trigger
time as time zero, $t_{0}$. In this fit the term $\exp(-t_{p}/t)$ was included
in the prompt function $f_{p}$ so that a peak position was found for the
prompt emission.  This peak time was then used as time zero
and a second fit done with $t_{p}=0$ (i.e. without an initial rise in the
prompt component).
Following this two stage procedure ensures that the prompt power law index
fitted, $\alpha_{p}$, is referenced with respect to the estimated peak time
rather than the somewhat arbitary BAT trigger time.
In most cases the time of the initial rise, $t_{a}$, of the afterglow
component, $f_{a}(t)$, was fixed at the transition time of the prompt
emission $t_{a}=T_{p}$. In a few cases this was shifted to later times
because a small dip was apparent in the light curves before the start
of the plateau or the plateau started particularly early.
There was no case in which the two components
were sufficiently well separated such that this time could be fitted as a free
parameter. i.e. we are unable to see the rise of the afterglow component
because the prompt component always dominates/persists at early times
and $t_{a}$ could be much less than $T_{p}$ for most GRBs.
Many of the decays exhibit flares towards the end of the prompt phase,
during the initial power law decay, on the plateau and even in the final
decay phase. All large flares were masked out of the fitting procedure.
Although apparently bright, such flares account for only $\sim10\%$ of the
total fluence in most cases.

Chi-squared fitting was performed in log(flux) vs. log(time) space
using the parameters $\log_{10}(T_{p})$,
$\log_{10}(T_{a})$ and logs of the products, i.e. $\log_{10}(F_{p}T_{p})$
and $\log_{10}(F_{a}T_{a})$.
The error estimation therefore produced a statistical error on the product of
flux and time directly and these products could then be used to 
calculate the fluence and an associated fluence error in each of the components.
The fluences of the prompt exponential and prompt powerlaw decay phases are
\begin{equation}
fl_{exp}=\frac{F_{p}T_{p}}{\alpha_{p}}(\exp(\alpha_{p})-1)
\label{eq3}
\end{equation}
\begin{equation}
fl_{dec}=\frac{F_{p}T_{p}}{\alpha_{p}-1}
\left(1-\left(\frac{T_{p}}{t_{max}}\right)^{\alpha_{p}-1}\right)
\label{eq4}
\end{equation}
where $t_{max}(>T_{p})$ is the end of the light curve or some late
time when the decay is deemed to have terminated. If $\alpha_{p}>>1$ then
$t_{max}$ can be set to infinity. The fluence of the exponential
phase in the afterglow component is reduced by the initial exponential rise
factor $\exp(-t_{a}/t)$ and is given approximately by
\begin{equation}
fl_{exp}=\frac{F_{a}T_{a}}{\alpha_{a}}
\left(\exp(\alpha_{a}\left(1-\frac{t_{a}}{T_{a}}\right))-1\right)
\end{equation}
The inclusion of the exponential rise term has negligible effect on the fluence
of the decay phase. Another way of viewing $\alpha_{p}$ (or
$\alpha_{a}$) is that
it controls the ratio of fluences seen from the exponential phase,
$t<T_{p}$ (or $T_{a}$), and the decay phase, $t>T_{p}$ (or $T_{a}$).
If the peak time,
$t_{p}$, is zero and $t_{max}\rightarrow \infty$ then the ratio
of the fluences for the prompt component is
\begin{equation}
\frac{fl_{exp}}{fl_{dec}}=\frac{(\exp(\alpha_{p})-1)(\alpha_{p}-1)}
{\alpha_{p}}
\label{eq6}
\end{equation}
and $fl_{exp}/fl_{dec}=1$ when $\alpha_{p}=\alpha_{1}\equiv 1.446$.
If $\alpha_{p}<<\alpha_{1}$ then the decay is slow and most of the
energy appears for $t>T_{p}$ in the power law decay.
If $\alpha_{p}>>\alpha_{1}$ then the decay is fast and most of the
energy appears for $t<T_{p}$ in the early exponential phase.
Fig. \ref{fig1} shows the fluence ratio as a function of $\alpha_{p}$.
A very similar expression holds for the fluence ratio of the
afterglow component but this
includes a minor adjustment because of the initial rise in the exponential
phase, $t<t_{a}$.

Table \ref{tab1} lists the fitted parameters for all the GRBs in the sample.
The type of decay fit (D) is also listed in Table \ref{tab3}.
Of the 107 decays fitted, 85 required two components in which the
afterglow component was dominant at the end of the observed light curve
(D=1).
In these cases the decay index of the prompt phase, $\alpha_{p}$ was usually
greater than the $\alpha_{a}$ the final decay index of the afterglow. In a
further 6 cases two components were required but the second
component appeared as a hump in the middle of the initial decay
and the prompt component reappeared and dominated again towards the end (D=2);
these are marked$^\ast$ in Table \ref{tab1}. For these objects the prompt
decay is slow, $\alpha_{p}<1.5$,
and it is usually the case that $\alpha_{p}<\alpha_{a}$.
The remaining 16 required only one
component (D=3) and did not exhibit a plateau phase with a subsequent power law
decay.
In 9 of these cases
this was probably because the object was faint and the prompt emission
faded below the XRT detection threshold before a putative plateau could be 
recognised.
Fig. \ref{fig2} shows typical examples of all these fitted types.

For 99 of the 107 GRBs the latter stages of the light curve are
well represented by the one or two component functional fit described
above. For those with two components the plateau gradually steepens in the
exponential phase, $t<T_{a}$, and relaxes into a simple power law
for $t\ge T_{a}$. For those with one component the prompt emission
turns over into a final power law decay at $t\ge T_{p}$.
However, in 8 cases there is clear evidence for a late temporal break. 
For these objects two extra
parameters were included in the fit, a final break at time
$T_{b}$ and a decay index $\alpha_{b}$ for $t>T_{b}$. Examples of these
are also shown in Fig. \ref{fig2}
and Table \ref{tab2} lists the fitted parameters.
Such a late break usually occurs in the final decay
of the afterglow component but in 3 GRBs, 060105, 060313 and 060607A,
the late break is seen in the power law decay of the prompt component.
For 060607A the break in the prompt component may have occured much
earlier and the final break could be coincident with the end of the
plateau, $T_{a}$ (see Fig. \ref{fig8}).
However, there is no doubt that a break
occurs near the end of the light curve and the decay after this break
is very steep.

Fig. \ref{fig3} shows the distribution of $T_{a}$ vs. $T_{p}$ for those
objects with two component fits.
There is no correlation between these times.\footnote{The error bars
plotted in Fig. \ref{fig3}, and all subsequent plots involving fitted values,
are 90\% confidence ranges.}
The frequency distributions of these times are shown in the bottom left panel.
The same figure shows the distribution of afterglow fluence
(the total fluence from the afterglow component) vs. the prompt fluence
calculated as the sum of the $T_{90}$ fluence seen by the BAT and the
fluence from the initial decay $fl_{dec}$ calculated from the XRT
flux using the equation above. The dotted line indicates those
objects for which the afterglow fluence is equal to the prompt fluence.
There are a few objects on or just above this line while the rest are
well below. There is a general trend that high prompt fluence leads
to high afterglow fluence, as might be expected, but the scatter about this
trend is large. This confirms the result from our earlier analysis
(O'Brien et al. 2006) but for a larger sample.
The frequency distributions of the fluences are shown in the lower right panel
of Fig. \ref{fig3}.

\section{Spectral evolution}

Spectral fitting with XSPEC (Arnaud 1996) version 11.3.2 
was used to determine the spectral index in the prompt phase
($\beta_{p}$, from the BAT data),
the prompt decay ($\beta_{pd}$ from the XRT data), on the plateau
($\beta_{a}$, XRT data) and in the final decay ($\beta_{ad}$, XRT data) for
$t>T_{a}$.
In some cases the coverage was poor and/or the count rate low so
it was not possible to separate $\beta_{pd}$ and $\beta_{a}$ or
$\beta_{a}$ and $\beta_{ad}$.
For the weakest bursts
it was only possible to derive $\beta_{p}$ and one spectral index
from the XRT, $\beta_{pd}$.
When fitting late time spectra the absorption was fixed to the early time
fitted values (both Galactic and intrinsic components) so that errors on 
the late time spectral indices were minimised.
Table \ref{tab3} lists these spectral indices for all the GRBs in the sample.
The $\pm$ ranges quoted are at 90\% confidence.
The decay fit type D is also listed.  When D=3 there is no 2nd component
and hence no plateau.  However, for some of these afterglows late time
XRT data are available and a late spectral index could be derived independently
from $\beta_{pd}$. These late spectral indicies are listed in the
$\beta_{ad}$ column.

Fig. \ref{fig4} shows the distribution of the afterglow plateau spectral
index $\beta_{a}$ vs. the prompt (BAT over $T_{90}$) spectral index
$\beta_{p}$. The range of indices from the prompt emission is large,
$-1.0$ to $2.2$ while the afterglow range is smaller, 0.4 to 2.3. Those
objects for which the prompt emission is especially soft ($\beta_{p}>1.8$) or
hard ($\beta_{p}<0.1$) evolve to produce an afterglow in the narrow
range $0.7<\beta_{a}<1.2$. The frequency distributions of all the
spectral indices are shown in bottom panels of Fig. \ref{fig4}.
The same figure shows the distribution of $\alpha_{a}$ vs.
$\alpha_{p}$ for decays $D=1$.
Again, there is no correlation but the range of decay
indices for the prompt component is large, 1.0 to 6.0, while the range for
the 2nd afterglow component is much smaller, 0.5 to 2.0 with one object at
$\alpha_{a}=2.7$. Note, there are 4 decays with $\alpha_{a}>3$ in 
Table \ref{tab1} but these are all type $D=2$ marked$^{\ast}$.
In the majority of objects $\alpha_{a}\le \alpha_{p}$.

\section{The expected coupling between $\alpha$ and $\beta$ in the afterglow
decay}

The standard fireball model predicts that there should be a simple
coupling between
the temporal decay index in the final afterglow, $\alpha_{a}$, and
the spectral index in the final afterglow, $\beta_{ad}$
(Sari, Piran \& Narayan 1998). The exact
form of the relationship depends on the density profile of the surrounding
circum-stellar medium (CSM) and whether the decay is observed
before or after a jet break  (e.g. Panaitescu \& Kumar 2002);
because the spectrum is being red-shifted as the jet is slowed down by
the CSM and after a jet break the peak flux of the observed synchrotron spectrum
is also decaying with time. The top left panel of Fig. \ref{fig5} shows
the temporal decay
index of the afterglow component, $\alpha_{a}$, vs. $\beta_{ad}$ the
spectral index in the final decay for those objects where $t_{max}>T_{a}$
and for which we have a significant measurement of $\beta_{ad}$ at $t>T_{a}$.
There is no correlation between
$\alpha$ and $\beta$ and if there is any trend at all it is that some
of the faster decays (large $\alpha$) occur for the smaller $\beta$ values.
The lower band shown indicates the region of a pre-jet break coupling
predicted by the model and the upper band the region expected
post-jet break. In each case the lower edge of the band corresponds to
the model in which the X-ray frequency is below the cooling break frequency
and the CSM density is constant. The upper edge corresponds to
the X-ray frequency above the cooling break frequency and a wind
density $\propto r^{-2}$.
Of the 70 objects plotted, 36 lie below the expectations
of the standard model in the bottom right of the plot. For 17
the upper limit of the 90\% confidence region in $\beta-\alpha$
doesn't intersect the pre-jet break band. So for $\approx 50\%$
of GRBs the spectral index of the afterglow is too large to produce
the observed temporal decay index. This can only occur under the model
if there is significant energy injection such that the peak of
the spectrum (or normalisation) is boosted fast enough to counteract the
drop expected from the change in red-shift as the outflow slows down.
It was pointed out by Nousek et al. (2006) that the spectrum and decay
of the plateau phase of several Swift GRBs was inconsistent with the
expectations of the model and that energy injection during this
phase was a possible explanation. The current 
analysis shows that the same is also true for many objects during
the subsequent decay phase, after the plateau. The lower left-hand panel
of Fig. \ref{fig5} shows $\alpha_{a}-(3/2)\beta_{ad}$ plotted as a function
of $T_{a}$. This function will be zero for pre-jet break afterglows
under the standard model 
(with uniform CSM and the X-ray frequency below the cooling frequency)
and negative for afterglows
with a value of $\alpha_{a}$ which is too small compared with the
$\beta_{ad}$. The horizontal dashed line is the pre-jet break expectation if
the X-ray frequency is above the cooling frequency. Consideration
of GRB efficiencies (Zhang et al. 2006) indicates that more than
60\% of afterglows in the sample described by O'Brien et al (2006) do
lie above the cooling break. If this is the case for the present, larger,
sample a significant fraction of the GRBs plotted as dots
(pre-jet break) are also  outside expectation.
The figure also indicates that there is no correlation of this function with
the time at which the final decay starts, so if energy injection is
the explanation it is occurring at large times, $>10^{5}$ seconds as well
as much earlier times.
Of the remaining $\sim50\%$ of objects plotted on the top left-panel
of Fig. \ref{fig5}, 26 lie within the pre-jet break
band while 8 lie above this close to or within the post-jet break
region. It is conceivable that some of the anomalously large
$\beta$ values could be associated
with the fitted absorption (both Galactic and intrinsic). We checked this
possibility but found no correlation between the position of an afterglow
on the $\beta-\alpha$ plane and the fitted $N_{H}$.

The top right-hand panel of Fig. \ref{fig5} shows the change in
spectral index into the final decay from the plateau,
$\Delta\beta=\beta_{ad}-\beta_{a}$ plotted vs. the plateau spectral index
$\beta_{a}$. Here there is a trend. If $\beta_{a}$ is small then
$\Delta\beta>0$ and the afterglow gets softer in the final decay. If
$\beta_{a}$ is large then $\Delta\beta<0$ and the final decay is harder. The 
scatter of index $\beta_{ad}$ is smaller in the final decay, being
confined to the range 0.0 to 2.2 as indicated on the top left plot.
The gradual narrowing of the spectral index range as the afterglows
develop can also be seen in the frequency distributions shown in the lower
panels of Fig. \ref{fig4}.
One object, GRB060218, has an anomously large final spectral index,
$\beta_{ad}=3$, but this GRB was very peculiar in many respects, in particular
for having a significant thermal component in the early X-ray spectrum,
Campana et al. (2006). No $\beta_{a}$ was derived for this afterglow because
the plateau is largely obscured by unusual, persistent, prompt emission.
The trend in the change in spectral index
from the plateau into the power law decay is independent of the position in
the $\beta-\alpha$ plane occupied by the final decay.

The above discussion has considered those objects with a 2nd afterglow
component that dominates in the later regions of the X-ray light curve (D=1).
22 of the X-ray decays required no 2nd component (D=3) or have a 2nd component
which produced a hump in the decay but faded towards the end (D=2).
In these cases the initial decay of the prompt component
dominates at late times.
The bottom right-hand panel of Fig. \ref{fig5} shows the correlation
between the late temporal decay index ($\alpha_{p}$) and spectral index
of these objects. The behaviour is very similar to objects (D=1) shown in 
the top left panel.
There is no obvious correlation and 8 of the 21 object fall below
the expected correlation in the bottom right. Also plotted
in this panel are the $\beta_{ad}$,$\alpha_{b}$ values for the late
breaks listed in Table \ref{tab2}. Five of these lie within the pre-jet
break band but three, GRB050814, GRB060105 and GRB060607A,
are within the post-jet break region
although the errors on the decay index after the
late break temporal break, $\alpha_{b}$, are rather large.
The complete X-ray light curve for GRB050814 containing the
late temporal break is shown in the upper right panel of Fig. \ref{fig8}.
Further analysis of GRB060105 is provided by Godet et al. (2006a).
GRB06067A is similar to GRB060105.

\section{Isotropic energy of the prompt and afterglow components}
\label{sece}

Fig. \ref{fig3} shows the correlation between the fluences
in the prompt component (including the BAT $T_{90}$ 15 to 150 keV and the
initial decay in the XRT 0.3 to 10 keV) and the afterglow component.
For those bursts for which we have a redshift (z values listed in Table
\ref{tab3})
these fluences can be used to estimate the equivalent isotropic energy.
We assumed a cosmology with $H_{o}=71$ km s$^{-1}$
Mpc$^{-1}$, $\Lambda=0.27$ and $\Omega=0.73$ and calculated $E_{iso}$ over
an energy band of 1 to 10000 keV in the rest frame applying a k-band correction
independently for the BAT and XRT data. For many objects we don't have
a measured value of the peak energy in the spectrum. In such cases we 
assumed $E_{peak}=116$ keV (the median value for Swift bursts)
and a spectral index of $\beta=2.3$ for $E>E_{peak}$.
The mean $E_{peak}$ for BATSE (pre-Swift) bursts was 235 keV (170-340 keV
90\% range) (Kaneko et al. 2006)
which is significantly larger than for {\em Swift} bursts, mean 138 keV,
median 116 keV.
However the mean redshift for {\em Swift} bursts considered here
is 2.46 while for
pre-{\em Swift} bursts it is $\approx1.52$ so we expect the mean/median
$E_{peak}$ for {\em Swift} bursts to be $\sim0.6$ the BATSE value.
The mean value for the spectral index above $E_{peak}$ for BATSE bursts
was 2.3 (Kaneko et al. 2006).
Fig. \ref{fig6} shows the equivalent isotropic energy 
for the two components. The symbols
used are the same as in Fig. \ref{fig5}. It is clear that the total
energy in the afterglow is not correlated with the position of the final
afterglow in the $\beta$-$\alpha$ plane.
Note that the correlation between the equivalent isotropic energy components
evident in Fig. \ref{fig6} is real but not particularly significant since
it arises from applying the measured redshift in both axes.
In many objects the total
energy seen in X-rays from the afterglow is significant compared with
the $\gamma$ and X-ray energy seen from the prompt component. The dotted
histogram in the right panel shows the distribution of $E_{iso}$ values
for GRBs observed by instruments pre-{\em Swift} taken from the tabulations in
Frail et al. (2001), Bloom et al. (2003) and Ghirlanda et al. (2004).
The maximum isotropic energy in the $Swift$ sample is similar to the
maximum seen previously, $\approx10^{54}$ ergs, but the distribution
of energies seen by $Swift$ is broader, has a lower mean
and extends to a lower limit of $\approx10^{47}$ ergs.

\section{Temporal breaks}

\begin{figure}[!htp]
\begin{center}
\includegraphics[height=19cm,angle=0]{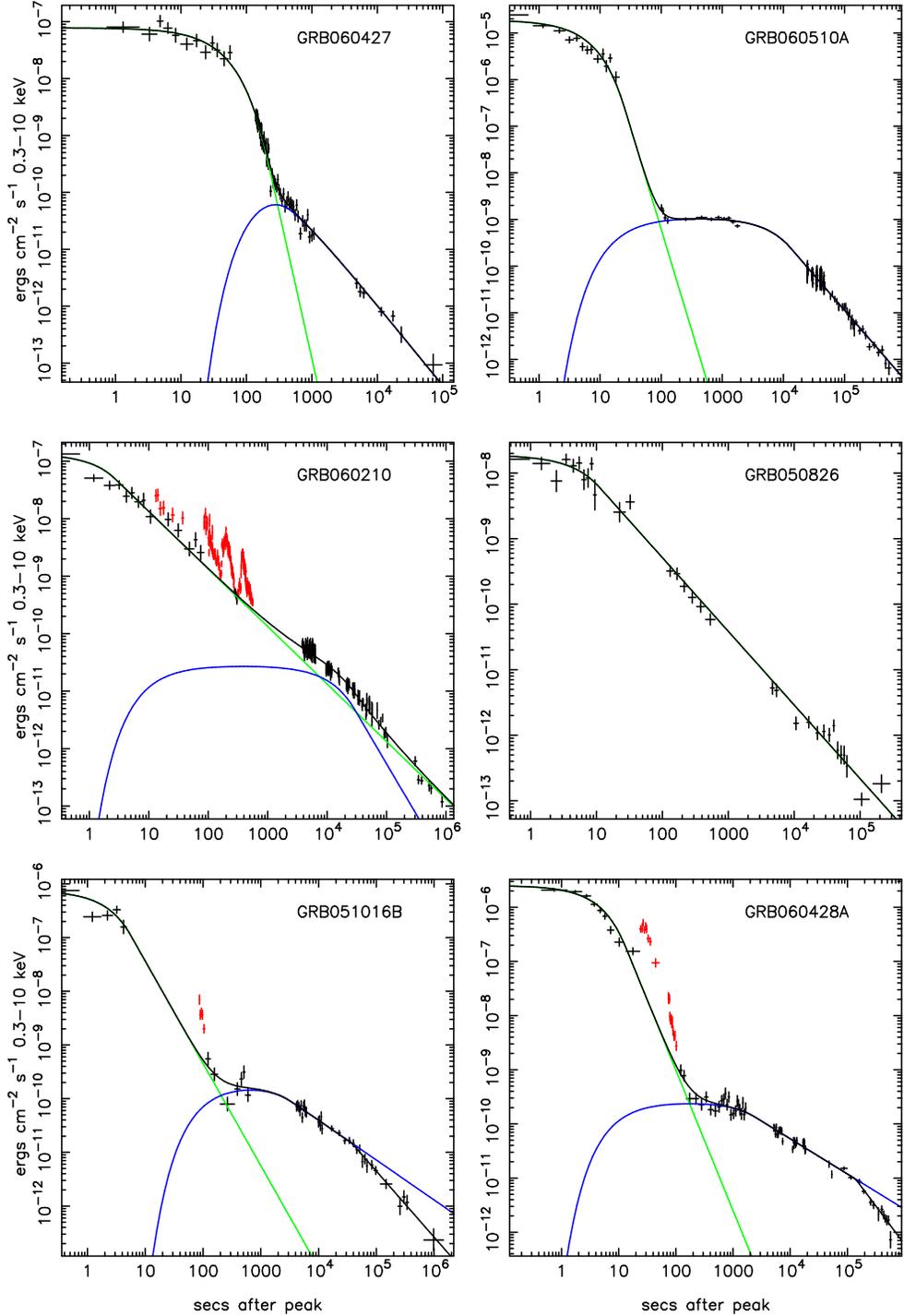}
\end{center}
\caption{Examples of the fits to X-ray decay curves. The prompt and
afterglow component functions are plotted in the same way as in
Fig. \ref{fig1}.
Top panels: the most
common type in which the 2nd, afterglow, component dominates at late times.
Middle panel left: a two component fit in which the afterglow component
forms a bump in the decay but the extrapolation of the
prompt component decay dominates at
late times. Middle panel right: a single component fit, requiring
no afterglow component. Bottom panels:
two examples of fits which include a late temporal break.
Flares are plotted in red and were excluded from the fitting procedure.}
\label{fig2}
\end{figure}

The visibility of the 2nd component used here to fit the X-ray decay curves
depends on the relative brightness of the prompt emission
decay compared with the afterglow plateau and the times $T_{p}$ and $T_{a}$.
If $T_{a}$ is not long after $T_{p}$ then the end of the plateau phase is not
visible, as is the case for the decay shown at the top left of Fig. \ref{fig2}.
However, for 64 of the 91 GRBs which required 2 components in the fit the
end of the plateau {\it is} visible, as is the case for the exemplar
GRB shown at the top right
of Fig. \ref{fig2}. In such cases the plateau gets slowly steeper
towards the end
of the exponential phase and eventually relaxes to a power law. There is
often no definitive or sharp break but the time $T_{a}$ is a robust
measure of where this transition occurs, taking into account all the data
available. Thus, the fitting provides 91 afterglow break times, $T_{a}$,
from a total of 107 objects.

In some cases it may be that any jet break time associated with
the edge of a putative jet becoming visible occurs at or before $T_{a}$.
In such decays we expect the subsequent afterglow to lie somewhere in the top
left of the $\beta-\alpha$ plane shown in Fig. \ref{fig5} and the 8 candidates
for such cases are shown as star symbols on this figure. Note that
the error bars shown in Fig. \ref{fig5} are at 90\% confidence and there
is one object lying in the lower pre-jet break band which is also
consistent with the upper post-jet break band.  The time $T_{a}$
for these GRBs is not a {\em pure} jet break time since in all cases the
$T_{a}$ marks the end of the plateau phase which does not behave
as a pre-jet afterglow (e.g. Nousek et al. 2006). The only evidence for a jet
break having occurred in these 8 candidates
is that the $\alpha$ and $\beta$ values of the
subsequent decay have the
right relationship for a post-jet break afterglow. Since 50\% of afterglows
don't agree with the expected alpha-beta value, including the pre-jet
break values, the argument in favour of jet breaks in these cases is weak.
However, $T_{a}$ is a reasonable
estimate of where any jet break may have occurred. Details of these
8 potential jet breaks are given at the top of Table \ref{tab4}.

As discussed above, an additional 8 GRBs required a late break to
fit the data as illustrated at the bottom of Fig. \ref{fig2}. The positions
of the final afterglows in the $\beta-\alpha$ plane for these GRBs are shown
in the bottom right of Fig. \ref{fig5} (plotted as triangles).
Only 4 of these objects are good
candidates for jet breaks and these are listed at the bottom of
Table \ref{tab4} and illustrated in Fig. \ref{fig8}.
The jet break time ranges plotted were estimated using the Ghirlanda
relation assuming $E_{peak}=116\pm50$ keV (there are no measured values
of $E_{peak}$ for these bursts). For GRB060607A, in the bottom
right panel, the late break occurs slighly earlier the allowed
band indicating that, for this break to be consistent with Ghirlanda,
the $E_{peak}$ should be $<66$ keV.
So, in summary, out of 72 afterglow breaks identified by
the fitting procedure only 12 are followed by an afterglow which is
consistent with post-jet break conditions and only 4 of these are
isolated breaks independent of the end of the plateau phase.
The remaining 60 have slow 
X-ray decays (low $\alpha$) and/or soft X-ray spectra (high $\beta$).

The interest in jet breaks stems from Equation \ref{eq1} which can provide
an estimate of the jet angle $\theta_{j}$ and hence the
collimation-corrected energy $E_{\gamma}$. For calculation of
all the subsequent $E_{\gamma}$ values considered below we have assumed
$n\eta_{\gamma}=0.6$ (or equivalently $n=3$ cm$^{-3}$ and
$\eta_{\gamma}=0.2$). $\eta_{\gamma}=0.2$ has
been widely assumed in pre-{\em Swift} analysis although recent work
using {\em Swift} data (Zhang et al. 2006) indicates that the efficiency
can be determined with more accuracy.
However, since $E_{\gamma}\propto (n\eta_{\gamma})^{1/4}$ the
resulting collimated energy is fairly insensitive to these parameters.
For 9 of the objects in
Table \ref{tab4} we have redshifts and can estimate $E_{\gamma}$ using
the jet break time. These are shown in Fig. \ref{fig7}.
The symbols are the same as in Fig. \ref{fig5}.
Also shown are the values derived for pre-{\em Swift}
GRBs using the tabulations in Frail et al. (2001), 
Bloom et al. (2003) and Ghirlanda et al. (2004).
For 31 others we have redshifts (as listed in
Table \ref{tab3}) but there is no break between $T_{a}$ and the last data
point at $t_{max}$. For these decays we can calculate a range of $E_{\gamma}$
which is excluded by the decay. These ranges
are plotted in the top right-hand panel
of Fig. \ref{fig7} with the energy corresponding to $T_{a}$ at the lower
end of each range and to $t_{max}$ at the upper end.
For many afterglows the excluded range of $E_{\gamma}$
covers a substantial fraction of the pre-{\em Swift} distribution.
The lower panels
of the same figure shows the respective frequency distributions. The 3 objects
with measured late breaks are in good agreement with the peak of the
pre-{\em Swift} $E_{\gamma}$ distribution published by Frail and Bloom.
Of the remaining 6 objects in which $T_{a}$ has been identified with
a possible jet break 2 lie within the lower wing of the pre-{\em Swift}
distribution and 4 lie below.
The distribution of $E_{\gamma}$ values derived from
the remaining $T_{a}$ values is similar in shape to the pre-{\em Swift}
distribution calculated using optically observed jet break times but
is offset to lower energies by a factor of $\approx34$. This corresponds to 
an average jet break time which is a factor $\approx110$ smaller or a jet angle
which is a factor $\approx10.5$ smaller.
The times $T_{a}$ derived from the X-ray decay curves
are, on average, a factor of $\approx80$ smaller than
the optical jet break times
observed for pre-{\em Swift} GRBs. The peak of the distribution of
$E_{\gamma}$ values calculated from $t_{max}$ is significantly higher than
the pre-{\em Swift} distribution because many of the X-ray decays extend
to later times without a temporal break.

$T_{a}$ is the only measure we have (or require) to specify the timescale
of the 2nd afterglow component. It is clear from the discussion above that,
in almost all cases, this does not represent a jet break time. However,
given that it is the only time extracted from the X-ray data it
may be related to the previously observed optical jet break times.
There are 14 GRBs for which we have a $T_{a}$ value, a redshift and a
measurement of the peak energy of the $\gamma$-ray spectrum, $E_{peak}$.
For these objects we can construct a Ghirlanda type correlation between the
peak energy in the rest frame $E_{peak}(z+1)$ and the collimated beam
energy $E_{\gamma}$ calculated assuming $t_{j}=T_{a}$.
This is shown alongside the conventional Ghirlanda
relation in Fig. \ref{fig9}. The X-ray measurements show a similar
behaviour, not as statistically significant as the optical version but
with about the same 
gradient and an offset in $E_\gamma$ by a factor of $\approx26$.
For these cases the X-ray break times, $T_{a}$, are a factor of $\approx90$ less
than pre-{\em Swift} optical break times.
Unfortunately none of the GRBs with
potential X-ray jet breaks identified and plotted in
Fig. \ref{fig8} has a measured $E_{peak}$ value so these can't be
included on Fig. \ref{fig9}.
However, we can calculate lower limits to $E_{\gamma}$, assuming 
$t_{j}=t_{max}$, for those decays without a late break but with an $E_{peak}$
and redshift measurement. These are included on Fig. \ref{fig9}.
They all lie to the right of the Ghirlanda correlation.
There is currently one afterglow observed by {\em Swift}, GRB050820A,
for which we have derived a time $T_{a}$ and the optical coverage extends to
$\sim100\times T_{a}$ (Cenko et al. 2006). In this case the optical decay
does appear to have a break at about the expected time,
$t_{j}\approx100\times T_{a}$. The Ghirlanda correlation was originally
formulated by considering the jet model and the collimated jet energy 
however, the correlation 
derived here, using $T_{a}$ from the X-ray afterglows, has more in common with
the model independent analysis described by Liang \& Zhang (2005).
The Liang-Zhang $E_{iso}-E_{peak}-t_{b,opt}$ correlation may well be
related to the $E_{iso}-E_{peak}-T_{a}$ correlation plotted in 
Fig. \ref{fig9}.

As a natural consequence of the above discussion, and
following Sato et al. (2006), we can adopt a different
approach and look for X-ray decays in which a predicted jet break is absent.
Apart from GRB050401 (or XRF050401, Mangano et al. 2006, Sakamoto et al. 2006),
GRB050416A and GRB050525A already considered by
Sato et al. (2006) there are 5 other X-ray decays monitored by Swift
which come under this category. Light curves for all 8 
are shown in Fig. \ref{fig10}.
In each case the expected range for the jet break time, $t_{j}$, is shown
as the shaded area. Each decay was followed for at least a factor of 100 times
longer than $T_{a}$ without any indication of a break in the temporal decay
index. For GRB060206 there is some indication that the X-ray light curve
is flattening for $t>10^{6}$ seconds. This is probably due to systematic
errors in the background subtraction when the afterglow is very faint and/or
contamination by a faint, nearby, background object.
GRB050525A, GRB050820A, GRB060510B and GRB060729 are cases where there
is a large flare at the end of the prompt phase and the late
BAT and/or early XRT data are not well fitted by the prompt function.
However, the plateau and subsequent
power law decay are well represented by the afterglow component in all
cases. The late
afterglow of  GRB060729 has currently been monitored for 77 days without
any indication of a break.

Optical data are available for 5 of the final power law decays of
these afterglows;
GRB050525A (Blustin et al. 2006),
GRB050820A (Aslan et al. 2006, Cenko et al. 2006),
GRB060206 (Stanek et al. 2006, Monfardini et al. 2006),
GRB060707 (de Ugarte Postigo et al. 2006, Jakobsson et al. 2006) and
GRB060729 (Grupe et al. 2006).
Optical jet breaks are not seen in any of these afterglows
although the decays of GRB050525A, GRB060206 do gradually steepen at late
times and there is a one final late measurement at $\sim3\times10^{6}$ seconds
for GRB050820A which indicates that the optical decay has turned down
(HST data presented in Cenko et al. 2006 as already mentioned above).
We have fitted these optical afterglows with a simple power law over
the period contemporary with the final X-ray decay ($t>T_{a}$) and the
results are shown in Table \ref{tab5}. For 4 afterglows the X-ray and
optical decay indices are consistent. For GRB050820A the optical index
is formally significantly lower than the X-ray index
but there is structure in the
optical decay curve which is not well modelled by a single power law fit.
For all 5 the X-ray and optical decays are remarkably similar and shallow with
decay indices in the range 0.75-1.5 and they all have $\alpha,\beta$ 
reasonably consistent with a pre-jet break afterglow.

Two other objects,
GRB050822 (Godet et al. 2006b) and GRB060218 (Campana et al. 2006)
also have {\em Swift} X-ray light curves with extended coverage of a
long power law
decay without a break. Both these objects have soft spectra and are
more reasonably considered as XRFs with $E_{peak}<25$ keV. For
GRB060218 we have a measured redshift of 0.0333 (Mirabal \& Halpern 2006).
Adopting a range of $E_{peak}$ values consistent with an XRF
(and a range of redshift values for GRB050822) it can be shown that
a jet break should have been seen at some point in the X-ray light-curves of
both these objects but none was observed (Godet et al. 2006b).
For GRB060218 radio observations have shown that the fireball was 
probably isotropic (Soderberg et al. 2006) and similarly, in the cases
of GRB050416A and 
GRB050822, it is possible that the jet opening angle was wide ($>20$ degrees).
For the XRFs a large opening angle and consequent absence of a jet break
is not unreasonable because $E_{iso}$ is typically rather low.
There are also 9 further GRBs (050318, 050603, 050802,
060108, 060115, 060418, 060502A, 060614 and 060714)
for which we can predict a range of jet break times and no break is seen.
For these objects the lowest predicted $t_{j}$ is well before the
last observed time but the upper limit on $t_{j}$ is after the
last observed time so we cannot rule out the possibility that
a break occurred too late (and too faint) to be seen by the {\em Swift} XRT.
There is a suggestion of a late break for GRB060614 but inclusion of such
a break is not statistically significant under the current analysis.

An important property of a jet break is that it should be achromatic,
occurring across the spectrum with the same temporal profile. Panaitescu et
al. (2006) show that temporal breaks for 6 GRBs seen by the {\em Swift} XRT
are not present in optical data which span the same period of
time. For 5 of these, GRB050802, GRB050922C, GRB050319, GRB050607
and GRB050713A the break in the X-ray decay is fitted as $T_{a}$ in the
analysis presented above. Therefore for these objects optical data do not
follow the X-ray profile modelled by the 2nd afterglow component and
the observed X-ray break is unlikely to be a jet break.
The remaining object, GRB050401, is fitted using 2 components but the
2nd component appears as a hump in the X-ray decay and the 1st component
has a slow decay, $\alpha_{p}=1.12\pm0.05$, that dominates near the end.
So in this case the optical data seem to follow
the behaviour of the power law decay of the 1st prompt component and
not the 2nd component. Similar behaviour is exhibited by
GRB060210 which has the same
relationship between the 1st and 2nd X-ray component, $\alpha_{p}=1.0\pm0.07$,
and the available optical data
(Stanek et al. 2006) follow the decay of the 1st component.
Conversely, there are examples of GRB decays
for which the X-ray and optical profiles do follow a similar pattern.
Optical data for GRB060206 (Stanek et al. 2006) have a profile which
closely follows the 2nd component of the X-rays. Stanek et al. argue
that for this GRB there is a break which can be seen simultaneously
in optical and X-rays but the present analysis finds no such X-ray break.
Instead there is a gentle curvature in the X-ray (and optical) decay which
is modelled by the slow transition from  exponential to power law in
the profile of the 2nd afterglow component. GRB050525A is another
example in which curvature in the later stages of the X-ray
and optical can be modelled as a break (Blustin et al. 2006). The
present analysis finds no such break in the later stages of the X-ray
decay of this burst either.

\section{Conclusions}

The X-ray decay curves of 107 GRBs observed by {\em Swift} have been
fitted in a systematic way using the simple functional form given
by Equation \ref{eq2}. They all require a prompt component
with parameters $T_{p}$, $F_{p}$ and $\alpha_{p}$ and 85 require
a 2nd afterglow component with parameters $T_{a}$, $F_{a}$ and $\alpha_{a}$.
The parameters and associated confidence limits 
are all listed in Table \ref{tab1}. $T_{p}$ is similar
to the familiar $T_{90}$ burst duration as discussed by O'Brien et al. (2006).
$\alpha_{p}$ is a prompt parameter which was
unavailable before the {\em Swift} era and indicates how fast the
prompt emission is decaying as also discussed by O'Brien et al. (2006).
The product $T_{p}F_{p}$ combined with $\alpha_{p}$ is a measure of the
prompt fluence (see Equations \ref{eq3} and \ref{eq4}).
$\alpha_{p}$ also determines the distribution of energy. If $\alpha_{p}<1.446$
then more energy is emitted for $t>T_{p}$ during the prompt
power law decay phase
and if $\alpha_{p}>1.446$
more energy is emitted during the prompt exponential phase $t<T_{p}$.
$T_{a}$ is the time when
the final afterglow power law decay starts, $\alpha_{a}$ is the index
of this final decay and also controls the curvature of the proceeding
plateau phase and the product $T_{a}F_{a}$ along with
$\alpha_{a}$ combine to give a fluence of the afterglow component (sum of
Equations \ref{eq3} and \ref{eq4} replacing subscripts $p$ with $a$).
$\alpha_{a}$ controls the distribution of energy in the 2nd functional
component in the same way as $\alpha_{p}$ controls this distribution in
the prompt component, as described above.
In addition, spectral fitting has provided $\gamma$ and X-ray spectral
indices over 
the phases of these functional fits as listed in Table \ref{tab3}.
The combination of Tables \ref{tab1} and \ref{tab3} provides a rich
data base for comparison with theoretical models.

The functional form used for the fits (Equation \ref{eq2})
is empirical rather than based on a physical model and it doesn't accomodate
flares. However, the exponential rise at time $t_{c}$ and transition
from exponential to power law at $T_{c}$ are reminiscent of the
theoretical discussion of the development of an afterglow given by
Sari (1997). Employing exponentials, rather
than power law sections with breaks, provides a curvature which neatly
fits the data and produces a remarkably good representation of the underlying
X-ray decay profile with the minimum number of parameters. In any
physical model the number of parameters that could influence the
shape of the X-ray light curves is large but some combination or subset
of these are likely to be represented by the values tabulated here.

Most ($\approx80\%$) GRBs have a 2nd afterglow component which dominates
at later times and this is probably the expected emission from the external
shock. For these objects the prompt and afterglow components appear
to be physically distinct.
Bursts marked$^{\ast}$ in Table \ref{tab1} (or D=2 in Table \ref{tab3})
have 2 component fits but
the 1st prompt component dominates at later times. For these
GRBs and those which only require 1 component in the fit (D=3 in 
Table \ref{tab3})
it is not clear where such an early prompt decay component comes from. It
could be the external shock but if this is the case then the external shock
is developing very early ($t<T_{p}$) and the prompt decay emission from this
shock is distinct from the more common external shock emission seen in
the 2nd component. In GRBs marked$^{\ast}$ (D=2 in Table \ref{tab3})
both these {\em external shock}
components are seen.  For these objects it is not so obvious that the prompt
and afterglow components fitted map directly to two physical entities.
The hump in the decay curve could be a signature of the presence of
different outflow structures (e.g. Eichler \& Granot 2006)
or a universal structured jet (e.g. M\'{e}sz\'{a}ros, Rees \& Wijers 1998).
Two of the GRBs, GRB051210 and GRB051221B, fitted by just one component (D=3 in Table \ref{tab3})
could be ``naked GRBs'' for which there is no external shock and the
prompt decay arises solely from high latitude emission (Kumar \& Panaitescu
2000). Such GRBs are expected to have $\alpha_{p}-\beta_{pd}=2$. This
difference is $1.81\pm0.2$ for GRB051210 and $1.73\pm1.0$ for GRB051221B
consistent with a high latitude interpretation.
Neither of these bursts were detected after the 1st orbit post trigger.
GRB051210 has an upper limit from the 2nd orbit consistent with the
steep power law decay. For GRB051221B upper limits were obtained from
the 2nd orbit and much later, $t=4.6\times10^{5}$ seconds, both again
consistent with a steep prompt decay and no afterglow or plateau.
The remaining 14 single component prompt decays were not steep with mean
$\alpha_{p}=1.35\pm0.22$
and significantly lower $\alpha_{p}-\beta_{pd}$ values
with a mean of $0.6\pm0.4$.
Although a rise time, $t_{a}$, was included in the afterglow function fit
in most cases
it was set at some arbitarily early time when any increase in this
component would be obscured by the prompt emission.
Thus, the functional fits provide very little evidence for any
observational signature of the ``onset'' of the afterglow or a rising
afterglow component predicted in some off-axis models 
(Eichler \& Granot 2006). The afterglow could be established very early.
Since the outflow is presumed to be moving close to the speed of light
there may be very little time delay between prompt emission and
emission from the external shock when the outflow
begins to decelerate.

\begin{figure}[!htp]
\includegraphics[height=15cm,angle=-90]{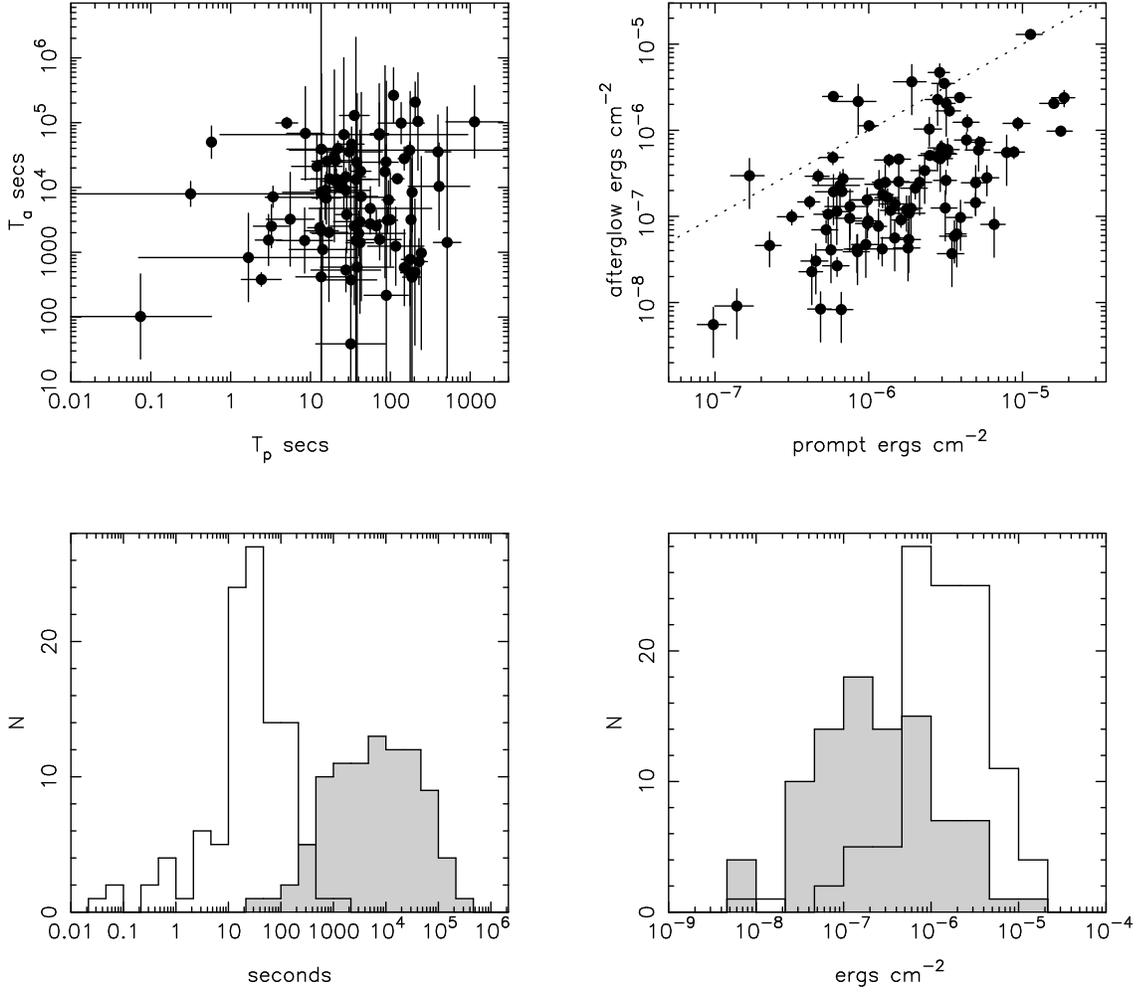}
\caption{Top left: The afterglow duration $T_{a}$ vs.
the prompt duration $T_{p}$ plotted for all objects in which the 2nd component
dominates the prompt component at late times.
Top right: The afterglow fluence calculated by integrating
the 2nd component in the XRT band 0.3-10 keV vs. the prompt fluence calculated
from the BAT $T_{90}$ flux 15-150 keV plus the XRT flux 0.3-10 keV in
the initial decay ($fl_{dec})$.
The dotted line indicates where the prompt and afterglow fluences are equal.
Bottom left: Frequency distributions of $T_{p}$ (open histogram)
and $T_{a}$ (shaded histogram).
Bottom right: Frequency distributions of prompt fluence (open histogram)
and afterglow fluence (shaded histogram).
}
\label{fig3}
\end{figure}

\begin{figure}[!htp]
\includegraphics[height=15cm,angle=-90]{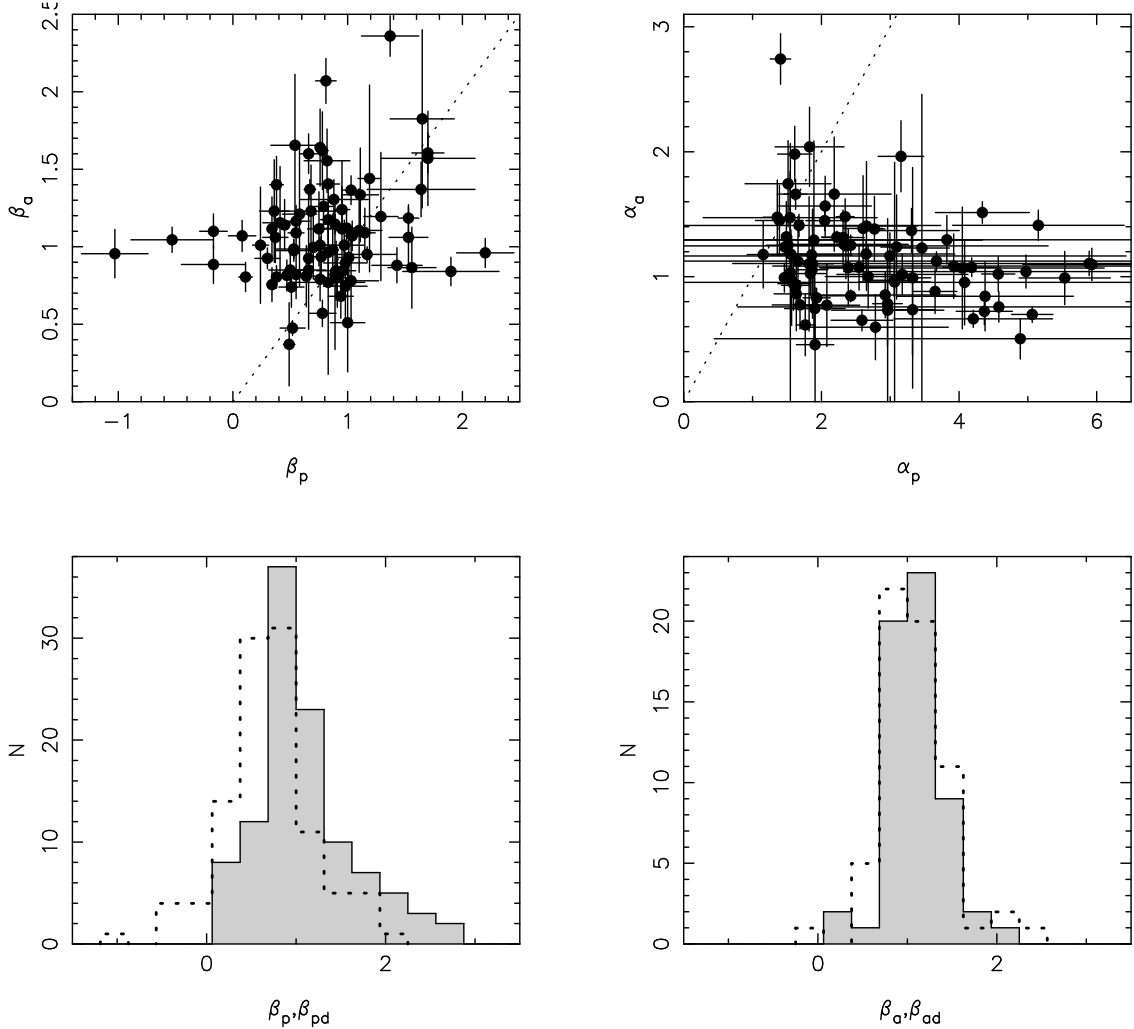}
\caption{Top left: Comparison of the spectral indices
$\beta_{a}$ (2nd, afterglow component) with $\beta_{p}$ (prompt BAT).
Top right: Comparison of the decay index for the 2nd component,
$\alpha_{a}$ with the prompt component, $\alpha_{p}$ for
decays type $D=1$. The dotted lines
indicate equality in both panels.
Bottom left: Frequency distributions of the prompt component spectral indices,
$\beta_{p}$ (dotted histogram) and $\beta_{pd}$ (shaded histogram).
Bottom right: Frequency distributions
of the afterglow component spectral indices,
$\beta_{a}$ (dotted histogram) and $\beta_{ad}$ (shaded histogram).
}
\label{fig4}
\end{figure}

There is no tight correlation between the prompt component parameters
and the afterglow component parameters. There is a trend that
the larger the prompt fluence the larger the afterglow fluence, as
might be expected, and the afterglow energy is usually considerably
less than the prompt energy. In some cases the afterglow energy is
about equal to the prompt but we never see afterglows which
are much more energetic than the prompt emission; see Figures \ref{fig3} and
\ref{fig6}.

Almost all of the X-ray afterglows end up in the same region of the 
$\beta-\alpha$ plane, $0.5<\beta<2$ and $0.4<\alpha<2$, independent of
the other parameters. However, $\approx50\%$ of these afterglows
are in the bottom right of the $\beta-\alpha$ plane with high $\beta$
and low $\alpha$, values not predicted by the standard fireball model.
This may be
due to persistent long lasting energy injection but this seems unlikely
in such a high percentage of objects and at such late times. Furthermore 
the trend expected from the
model, that high $\beta$ values should give high $\alpha$ values 
and vice versa, is not
seen at any level of significance. If this is because of energy injection
this energy injection is occurring in just the right number of afterglows
and at just the right level such that the expected $\beta-\alpha$ coupling is
hidden or masked. If, for example, energy injection were always present at
some fixed level all afterglows would be shifted to lower $\alpha$ values
but any correlation would be preserved. In fact, if energy injection
occurred preferentially at late times post-jet break afterglows would
be shifted to smaller $\alpha$ values but pre-jet break afterglows
would not be altered and the overall correlation
including both pre and post jet break afterglows might be tightened.
The curvature or break
which marks the end of the afterglow plateau phase at time $T_{a}$ is
often accompanied by a small spectral change such that soft afterglows
become somewhat harder and hard afterglows become softer bringing the
final afterglow spectral index into the rather narrow range indicated above.

\begin{figure}[!htp]
\includegraphics[height=15cm,angle=-90]{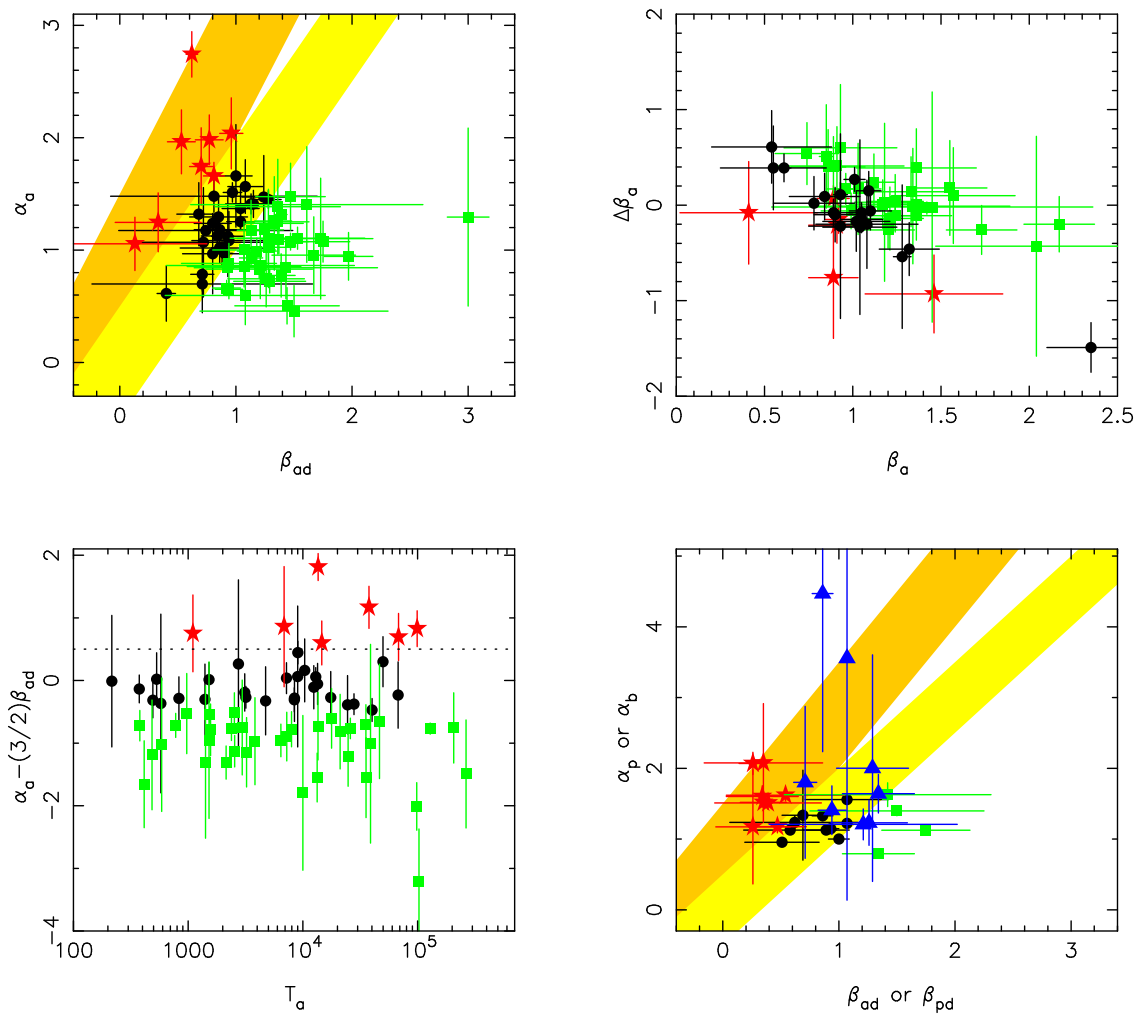}
\caption{Top left-hand panel: the temporal decay index of the afterglow
plotted vs. the spectral decay index in the final afterglow decay. The lower
band (yellow) indicates the area expected to be occupied by a pre-jet break
afterglow and the upper band (orange) the area post-jet break.
GRBs which fall in the pre-jet break region are plotted as circular dots
(black), those which fall above this in the post-jet break region are
plotted as stars (red) and those below the pre-jet break band are plotted
as squares (green). The same symbols are used in the other three panels.
Top right-hand panel: change in spectral index $\beta_{ad}-\beta_{a}$
vs. the plateau spectral index $\beta_{a}$.
Bottom left-hand panel: $\alpha_{a}-(3/2)\beta_{ad}$ vs. the duration of
the afterglow $T_{a}$. This function is expected to be zero for 
a pre-jet break afterglow in a uniform CSM with the X-ray frequency
below the cooling break frequency. The dotted line indicates the zero
if the X-ray frequency is above the cooling break frequency.
Bottom right-hand panel: the temporal decay index vs. the spectral index
for afterglow power law decay in objects which require only one component
in the fitting or in which the prompt component dominates at the end of
the X-ray light-curve.
The $\beta_{ad}$ vs. $\alpha_{b}$ for the late temporal breaks listed
in Table \ref{tab2} are also shown as triangles (blue).
The bands shown are the same as for the top left-hand
panel.}
\label{fig5}
\end{figure}

\begin{figure}[!htp]
\includegraphics[height=15cm,angle=-90]{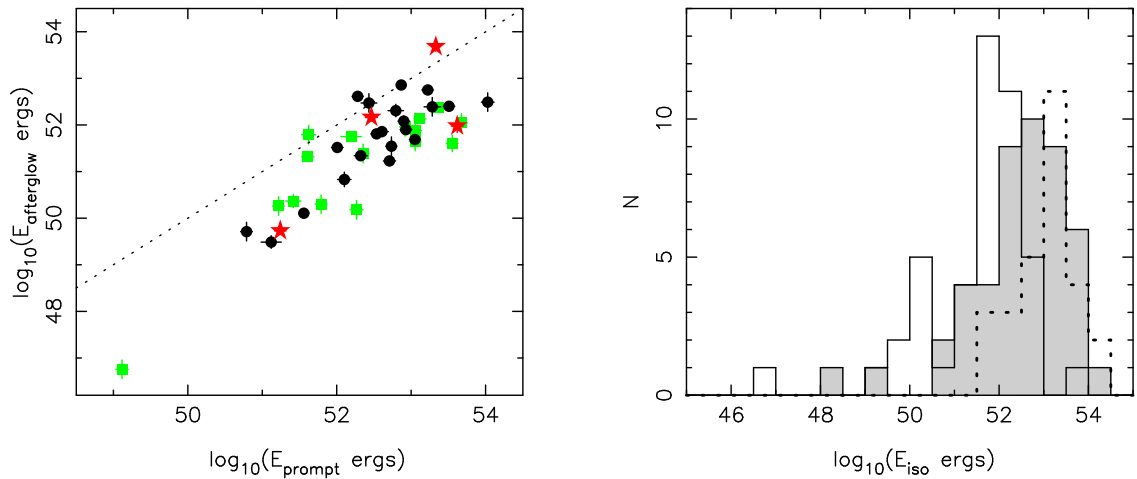}
\caption{Left: The isotropic afterglow energy vs. the isotropic prompt
energy. Symbols indicate the position of the afterglow in the $\beta$-$\alpha$
plane as in Fig. \ref{fig5}. The dotted line indicates equality between
the prompt and afterglow energies.
Right: The distribution of isotropic energies.
Filled histogram $E_{prompt}$ and solid line histogram $E_{afterglow}$.
The dotted histogram is the distribution of $E_{iso}$ for
GRBs observed before the {\em Swift} era as tabulated by Frail et al. (2001),
Bloom et al. (2003) and Ghirlanda et al. (2004).}
\label{fig6}
\end{figure}

\begin{figure}[!htp]
\includegraphics[height=15cm,angle=-90]{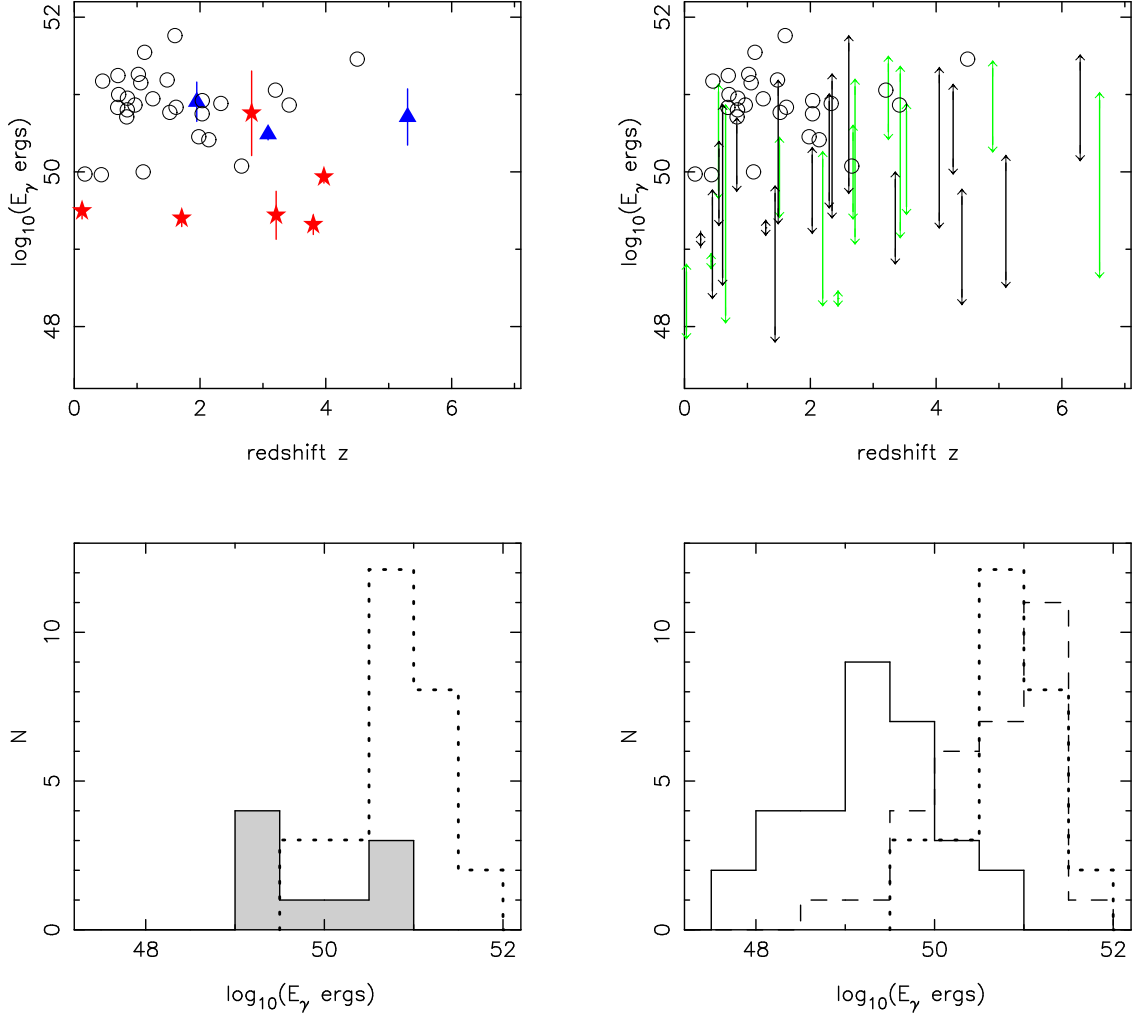}
\caption{The jet energy $E_{\gamma}$ calculated using equation \ref{eq1}.
Top left: The stars and triangles correspond to the jet break times listed in
Table \ref{tab4} (symbols as Fig. \ref{fig5}).
The open circles are the values obtained using the $t_{j}$ and 
$E_{iso}$ tabulated by Frail et al. (2001),
Bloom et al. (2003) and Ghirlanda et al. (2004).
Top right: Ranges of $E_{\gamma}$ excluded by the X-ray decays without 
breaks, $t>T_{a}$. The lower points are from $T_{a}$ and upper
from $t_{max}$ (the end of the light curve).
Bottom left: The filled histogram shows the distribution of $E_{\gamma}$
for the breaks listed in Table \ref{tab4}.
The dotted histogram is the Frail-Bloom-Ghirlanda sample.}
Bottom right: The dotted histogram is the Frail-Bloom-Ghirlanda sample.
The solid line histogram is the distibution derived from $T_{a}$
and the dashed line histogram from $t_{max}$.
\label{fig7}
\end{figure}

\begin{figure}[!htp]
\includegraphics[height=13cm,angle=0]{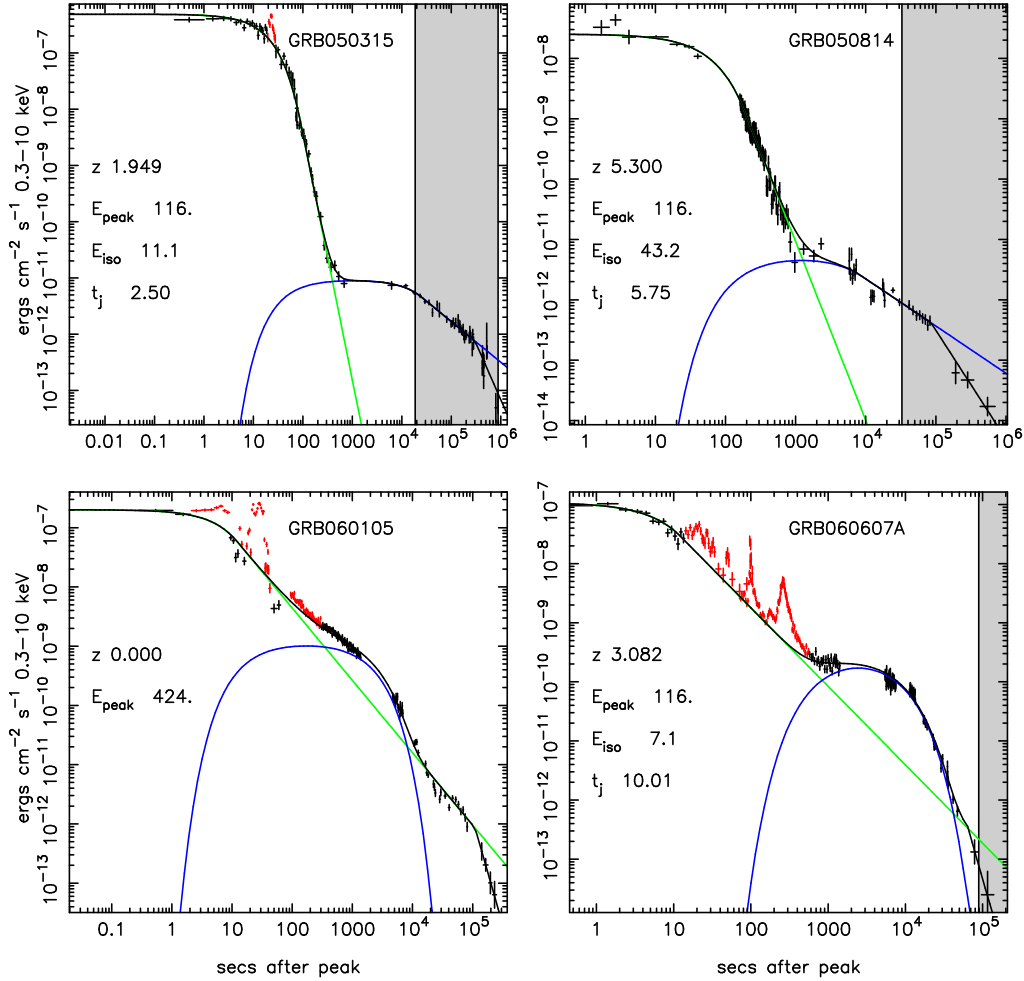}
\caption{X-ray decays in which there is potential jet break.
The shaded area indicates the expected range of $t_{j}$ calculated
using the Ghirlanda relation for
GRBs with a measured redshift. None of these has
a measured $E_{peak}$ value so 
$116\pm50$ keV has been assumed (see Section \ref{sece}).
The estimated $E_{iso}$ ($10^{52}$ ergs) and
predicted $t_{j}$ (days) are listed on the left.}
\label{fig8}
\end{figure}

\begin{figure}[!htp]
\includegraphics[height=7cm,angle=-90]{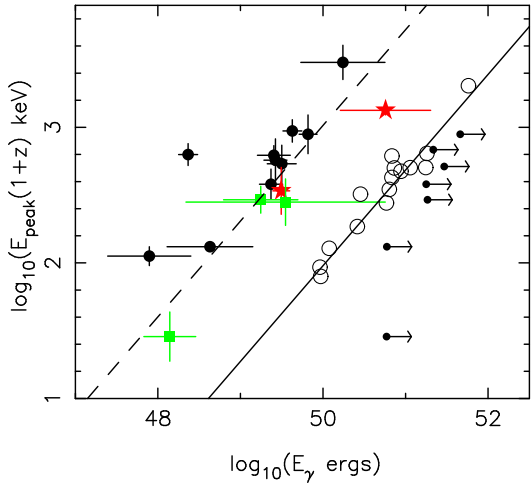}
\caption{The open circles and solid line are the Ghirlanda relation.
The points with error bars are {\em Swift} GRBs for which
values of $E_{\gamma}$ were derived using the Ghirlanda relation
assuming $t_{j}=T_{a}$.
The symbols plotted are the same as in Fig. \ref{fig5}.  The dashed
line is a fit to the {\em Swift} sample assuming the slope is fixed
at 0.706, the best fit value from Ghirlanda et al. (2004).
The lower limits of energy shown were derived using the latest
observed time as $t_{j}$ for GRBs with lasting afterglow decays which show no
break (the light curves plotted in Fig. \ref{fig10}).
$E_{peak}$ values used here and for Figures \ref{fig8} and \ref{fig10}
were taken from Cummings et al. (2005), Crew et al. (2005),
Golenetskii et al. (2005, 2006) or they were derived from spectal
fitting of the BAT data.}
\label{fig9}
\end{figure}

For the 91 GRBs which require a 2nd afterglow component we derive time
$T_{a}$ which marks the end of the plateau and the start of the final decay.
For 64 cases we can see this transition/break in the data.
Do we see any jet breaks? For 8 GRBs
the final temporal decay and spectral index $t>T_{a}$ are consistent
with the jet model after a jet break has occurred but for 6 of these objects
for which we have a redshift the $E_{\gamma}$ values are low compared
with the distribution derived from consideration of optical jet breaks and
the implied jet angles are small. 8 X-ray decays have a late break,
$T_{b}>T_{a}$,
and 4 of these have final $\alpha$ and $\beta$ values consistent with
a post jet break afterglow. 3 of these have redshifts and $E_{\gamma}$
values which are in accord with the pre-{\em Swift} distribution.
On the other hand
there are 11 decays with long-lasting coverage, $t>T_{a}$, in which
a jet break predicted using the Ghirlanda relation is definitely not seen
and there are a further 9 objects in which a break is not seen but
it may have occurred at a rather late time beyond the coverage provided
by the sensitivity limit of the {\em Swift} XRT.
13 decays (in the top right and bottom left panels
of Fig. \ref{fig5}) have been identified as lying in the post jet-break
region of the $\beta-\alpha$ plane (star symbols). For such decays the electron
energy distribution index is expected to be equal to the decay index.
Two of these decays, GRB060421 and GRB060526, have 
decay indices $<1.6$ (see Table \ref{tab4}), probably too low to be
the same as the electron index.

There are currently $\sim10$ afterglows for which there are both
X-ray and optical data available during the X-ray plateau and into the final
decay (and there are a few more objects for which such data will soon be
publically available). The X-ray and optical often follow a similar trend but
temporal breaks which are seen in one band are not always seen in the other.
This is probably because the end of the plateau and start of the final power
law is marked by a continuous curvature such that the light curves
in both bands are slowly getting steeper. Rather than fitting isolated
temporal breaks independently in either band we suggest that
simultaneous fitting of a profile as described by Equation \ref{eq2}
in an attempt to find a value for $T_{a}$ which is consistent
with both the X-ray and optical data would be more illuminating. Such
a joint fit would also provide a best estimate of the fluence ratio
between the X-ray and optical bands (or equivalently the X-ray/optical
flux ratio).

If we assume that $T_{a}$ is a jet break time
$t_{j}$ and calculate $E_{\gamma}$ for those objects with redshifts the
distribution of $E_{\gamma}$ which results is very similar to the 
distribution derived from optical jet breaks but is offset to lower energies
by a factor $\approx34$ implying that any optical jet break should be seen
at time $t_{j}\approx110\times T_{a}$. Furthermore, for those GRBs for which
we have an $E_{peak}$ value for the $\gamma$-ray spectrum
the peak energy in the rest frame is correlated with the collimated
energy estimate $E_{\gamma}$, in the form of a Ghirlanda relation,
but with the $E_{\gamma}$ values offset to lower energy by a factor $\approx26$
which implies that $t_{j}\approx90\times T_{a}$. It appears that
$T_{a}$ extracted from X-ray decays in the present analysis
has properties related to $t_{j}$ derived from optical data.
Whether this apparent connection is purely statistical in nature or
has some deeper significance remains to be seen but it is doubtful
that either $T_{a}$ or $t_{j}$ are actually ``jet break'' times.
However, it is likely that the end of the plateau phase, $T_{a}$, does
depend on the total energy in the outflow, the collimation angle
of the outflow and the density of the CSM and that the
$E_{iso}-E_{peak}-T_{a}$ correlation reported here is related to
the $E_{iso}-E_{peak}-t_{b,opt}$ relation discussed by
Liang \& Zhang (2005).

\acknowledgments

The authors would like to thank Alin Panaitescu for
useful comments and feedback on early drafts of the paper
and we gratefully acknowledge funding for {\em Swift} at the
University of Leicester by PPARC, in the USA by NASA and in Italy
by ASI.





\clearpage



\begin{figure}[!htp]
\includegraphics[height=13cm,angle=0]{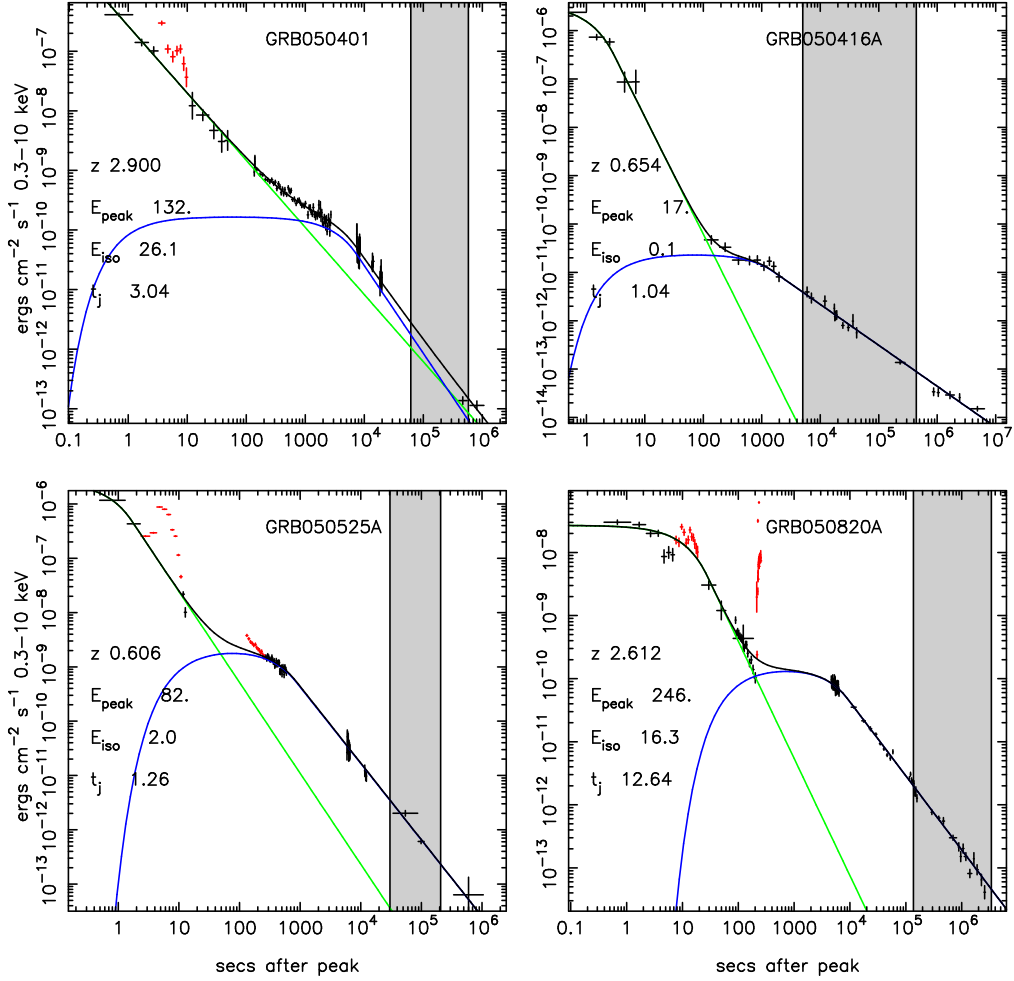}
\caption{X-ray decays in which a predicted jet break is absent.
$E_{peak}$ listed in keV, $E_{iso}$ in 10$^{52}$ ergs and $t_{j}$ in days.
The shaded area indicates the expected range for $t_{j}$. For
GRB060510B and GRB060729 there is no measured $E_{peak}$ value so 
$116\pm50$ keV has been assumed (see Section \ref{sece}).}
\label{fig10}
\end{figure}
\addtocounter{figure}{-1}
\begin{figure}[!htp]
\includegraphics[height=13cm,angle=0]{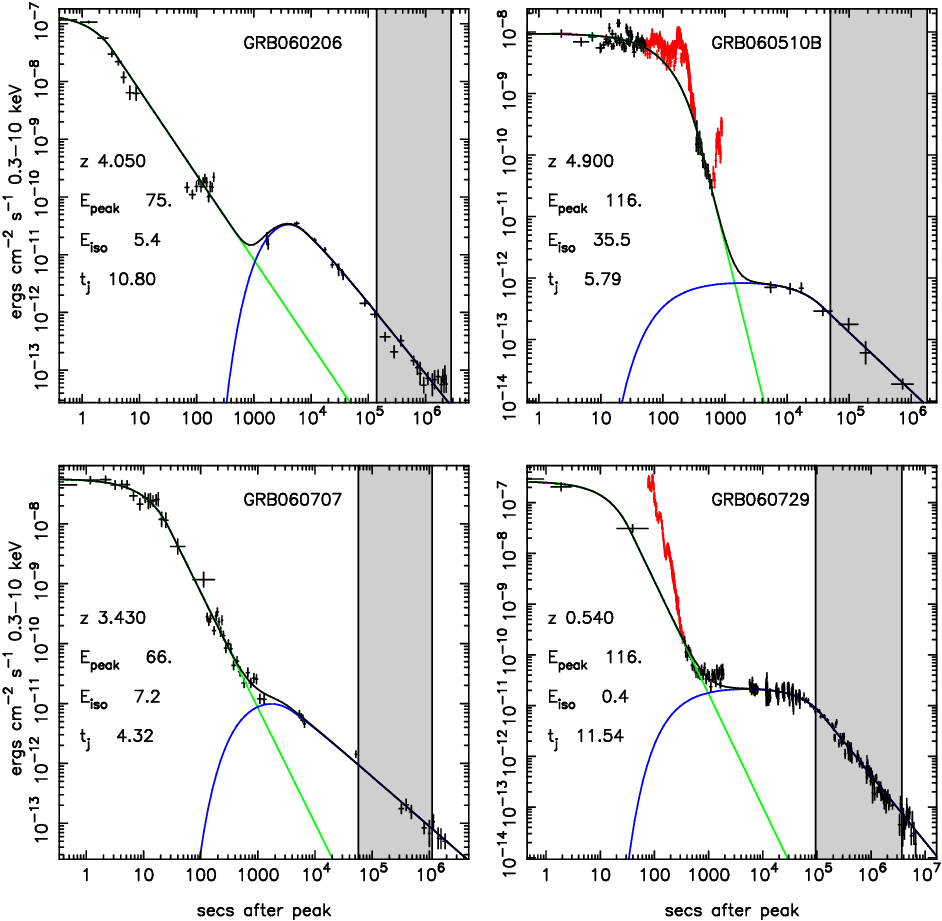}
\caption{cont.}
\end{figure}

\clearpage

\input{tab1}
\input{tab2}
\input{tab3}
\input{tab4}
\input{tab5}

\end{document}

%% file: tab1.tex
 \begin{deluxetable}{lrrrrrrr}
 \tablecaption{
 Best fit parameters for the Swift X-ray decay curves
 together with upper and lower
 90\% confidence limits.
 Times are in seconds and fluxes in
 ergs cm$^{-2}$ s$^{-1}$ 0.3-10 keV.
 $t_{p}$ is the prompt peak time with respect
 to the BAT trigger time used as time zero.
 $^{\ast}$2 component fits in
 which the 1st prompt component
 dominates towards the end of the X-ray light curve.
 $^{\dagger}$These GRBs have less than 50\% of $T_{90}$
 included in the fitting.\label{tab1}}
 \tablehead{
 \colhead{GRB}&\colhead{$t_{p}$}&\colhead{$\log_{10}(T_{p})$}&
 \colhead{$\log_{10}(F_{p}T_{p})$}&\colhead{$\alpha_{p}$}&
 \colhead{$\log_{10}(T_{a})$}&\colhead{$\log_{10}(F_{a}T_{a})$}&
 \colhead{$\alpha_{a}$}}
 \startdata
 050126&
 $  0.89$&
 $  1.95_{\;  1.70}^{\;  2.26}$&
 $ -6.96_{ -8.10}^{ -6.86}$&
 $  4.05_{\;  3.04}^{\;  7.17}$&
 $  2.34_{\;  1.34}^{\;  5.64}$&
 $ -8.10_{-10.74}^{ -6.47}$&
 $  1.07_{\;  0.63}^{\;  1.61}$
 \\
 050128&
 $  1.50$&
 $  0.51_{\;  0.32}^{\;  0.77}$&
 $ -7.02_{ -7.11}^{ -6.89}$&
 $  1.16_{\;  1.02}^{\;  1.49}$&
 $  3.40_{\;  3.07}^{\;  3.73}$&
 $ -6.88_{ -7.10}^{ -6.79}$&
 $  1.18_{\;  0.99}^{\;  1.52}$
 \\
 050219A$^{\dagger}$&
 $  0.50$&
 $  1.30_{\;  0.30}^{\;  2.30}$&
 $ -7.07_{ -7.12}^{ -7.03}$&
 $  1.55_{\;  1.49}^{\;  1.61}$&
 $  4.49_{\;  3.49}^{\;  5.81}$&
 $ -7.28_{ -8.89}^{ -6.03}$&
 $  1.03_{\;  0.00}^{\; 11.63}$
 \\
 050315&
 $  2.04$&
 $  1.95_{\;  1.89}^{\;  2.01}$&
 $ -6.26_{ -6.43}^{ -6.16}$&
 $  4.37_{\;  4.03}^{\;  4.84}$&
 $  4.39_{\;  4.13}^{\;  4.66}$&
 $ -6.94_{ -7.07}^{ -6.82}$&
 $  0.72_{\;  0.63}^{\;  0.84}$
 \\
 050318&
 $  0.00$&
 $ -1.13_{ -2.13}^{ -1.22}$&
 $ -6.59_{ -8.40}^{ -4.23}$&
 $  1.39_{\;  1.26}^{\;  1.96}$&
 $  2.01_{\;  1.37}^{\;  2.68}$&
 $ -5.99_{ -6.30}^{ -5.77}$&
 $  1.45_{\;  1.22}^{\;  1.78}$
 \\
 050319$^{\dagger}$&
 $ 10.56$&
 $  1.52_{\;  1.37}^{\;  1.68}$&
 $ -5.87_{ -5.94}^{ -5.80}$&
 $  2.61_{\;  2.37}^{\;  2.94}$&
 $  4.67_{\;  4.40}^{\;  4.93}$&
 $ -6.70_{ -6.78}^{ -6.61}$&
 $  1.39_{\;  1.01}^{\;  1.86}$
 \\
 050401$^{\ast}$&
 $  0.02$&
 $ -0.31_{ -1.31}^{\;  0.07}$&
 $ -6.54_{ -6.61}^{ -6.05}$&
 $  1.13_{\;  1.09}^{\;  1.18}$&
 $  3.87_{\;  3.69}^{\;  4.11}$&
 $ -6.54_{ -6.62}^{ -6.48}$&
 $  1.47_{\;  1.27}^{\;  1.94}$
 \\
 050406&
 $  0.62$&
 $  1.13_{\;  0.73}^{\;  1.37}$&
 $ -6.90_{ -7.11}^{ -6.66}$&
 $  3.06_{\;  2.39}^{\;  3.42}$&
 $  2.62_{\;  1.62}^{\; 12.22}$&
 $ -8.42_{-14.25}^{\:\:  7.36}$&
 $  0.96_{\;  0.00}^{\;  5.04}$
 \\
 050412$^{\ast}$&
 $  0.90$&
 $  0.90_{\;  0.64}^{\;  1.43}$&
 $ -7.95_{ -8.12}^{ -7.75}$&
 $  1.17_{\;  0.83}^{\;  2.45}$&
 $  2.72_{\;  0.30}^{\;  5.29}$&
 $ -9.05_{-67.12}^{ -7.05}$&
 $  5.63_{\;  0.00}^{\; 11.27}$
 \\
 050416A&
 $  0.04$&
 $  0.47_{\;  0.32}^{\;  0.66}$&
 $ -6.03_{ -6.21}^{ -5.95}$&
 $  2.43_{\;  2.17}^{\;  3.11}$&
 $  3.19_{\;  2.69}^{\;  3.48}$&
 $ -7.79_{ -7.94}^{ -7.66}$&
 $  0.85_{\;  0.80}^{\;  0.90}$
 \\
 050421$^{\ast}$&
 $  0.00$&
 $  1.12_{\;  0.48}^{\;  1.83}$&
 $ -7.54_{ -7.86}^{ -7.19}$&
 $  1.34_{\;  1.07}^{\;  2.34}$&
 $  2.32_{\;  2.11}^{\;  2.88}$&
 $ -7.24_{ -7.36}^{ -6.59}$&
 $  4.79_{\;  2.63}^{\; 10.06}$
 \\
 050422&
 $  1.99$&
 $  1.58_{\;  1.25}^{\;  2.17}$&
 $ -6.79_{ -7.14}^{ -6.63}$&
 $  2.78_{\;  2.33}^{\;  4.45}$&
 $  2.77_{\;  1.77}^{\;  5.45}$&
 $ -9.04_{-11.87}^{ -7.71}$&
 $  0.60_{\;  0.32}^{\;  0.84}$
 \\
 050502B&
 $  0.53$&
 $  1.50_{\;  0.86}^{\;  1.74}$&
 $ -7.98_{ -8.65}^{ -7.67}$&
 $  4.58_{\;  3.62}^{\; 11.21}$&
 $  1.59_{\;  0.59}^{\;  4.35}$&
 $ -8.51_{-10.20}^{ -8.10}$&
 $  0.76_{\;  0.61}^{\;  0.86}$
 \\
 050505&
 $  3.18$&
 $  1.58_{\;  1.41}^{\;  2.45}$&
 $ -6.44_{ -7.36}^{ -6.33}$&
 $  1.55_{\;  1.37}^{\;  3.90}$&
 $  4.39_{\;  4.15}^{\;  4.87}$&
 $ -6.92_{ -7.28}^{ -6.87}$&
 $  1.47_{\;  1.31}^{\;  2.05}$
 \\
 050509B&
 $  0.01$&
 $ -1.52_{ -1.78}^{ -1.20}$&
 $ -9.24_{ -9.43}^{ -9.02}$&
 $  1.17_{\;  1.09}^{\;  1.27}$&
 \\
 050525A$^{\dagger}$&
 $  0.60$&
 $  0.22_{ -0.78}^{ -0.25}$&
 $ -6.09_{ -8.09}^{ -4.09}$&
 $  1.67_{\;  0.00}^{\;  4.12}$&
 $  2.92_{\;  1.69}^{\;  3.06}$&
 $ -6.33_{ -6.39}^{ -6.31}$&
 $  1.41_{\;  1.30}^{\;  1.48}$
 \\
 050603&
 $  0.23$&
 $  0.94_{\;  0.83}^{\;  1.29}$&
 $ -7.13_{ -7.32}^{ -7.08}$&
 $  1.52_{\;  1.41}^{\;  2.66}$&
 $  4.83_{\;  3.69}^{\;  5.13}$&
 $ -7.12_{ -7.36}^{ -7.08}$&
 $  1.74_{\;  1.52}^{\;  2.21}$
 \\
 050607&
 $  0.49$&
 $  1.08_{\;  0.83}^{\;  1.46}$&
 $ -6.63_{ -6.73}^{ -6.47}$&
 $  1.87_{\;  1.70}^{\;  2.30}$&
 $  4.32_{\;  3.80}^{\;  4.81}$&
 $ -7.96_{ -8.18}^{ -7.85}$&
 $  1.10_{\;  0.79}^{\;  1.66}$
 \\
 050712&
 $  3.55$&
 $  1.86_{\;  1.74}^{\;  1.98}$&
 $ -6.60_{ -6.65}^{ -6.55}$&
 $  1.57_{\;  1.46}^{\;  1.70}$&
 $  4.83_{\;  4.30}^{\;  5.24}$&
 $ -7.47_{ -7.71}^{ -7.34}$&
 $  0.97_{\;  0.65}^{\;  1.36}$
 \\
 050713A$^{\dagger}$&
 $  1.00$&
 $  1.31_{\;  1.21}^{\;  1.57}$&
 $ -6.17_{ -6.73}^{ -6.09}$&
 $  2.65_{\;  2.22}^{\;  4.80}$&
 $  4.09_{\;  3.90}^{\;  4.29}$&
 $ -6.81_{ -6.86}^{ -6.75}$&
 $  1.18_{\;  1.09}^{\;  1.30}$
 \\
 050713B&
 $  0.22$&
 $  2.17_{\;  2.09}^{\;  2.24}$&
 $ -6.43_{ -6.54}^{ -6.34}$&
 $  3.06_{\;  2.74}^{\;  3.39}$&
 $  4.45_{\;  4.27}^{\;  4.63}$&
 $ -6.57_{ -6.66}^{ -6.49}$&
 $  0.97_{\;  0.85}^{\;  1.11}$
 \\
 050714B&
 $ 32.63$&
 $  2.27_{\;  2.15}^{\;  2.39}$&
 $ -6.43_{ -6.88}^{ -6.36}$&
 $  4.89_{\;  4.51}^{\; 13.41}$&
 $  2.62_{\;  1.62}^{\;  4.44}$&
 $ -7.84_{-12.36}^{ -7.28}$&
 $  0.50_{\;  0.34}^{\;  0.66}$
 \\
 050716$^{\dagger}$&
 $  0.35$&
 $  1.93_{\;  1.88}^{\;  1.99}$&
 $ -6.38_{ -6.41}^{ -6.35}$&
 $  1.66_{\;  1.60}^{\;  1.73}$&
 $  4.24_{\;  3.24}^{\;  5.88}$&
 $ -7.51_{ -8.56}^{ -7.07}$&
 $  1.12_{\;  0.75}^{\;  1.47}$
 \\
 050717&
 $  1.47$&
 $  1.36_{\;  1.33}^{\;  1.39}$&
 $ -6.34_{ -6.35}^{ -6.33}$&
 $  1.52_{\;  1.49}^{\;  1.55}$&
 \\
 050721&
 $  0.14$&
 $  1.57_{\;  1.46}^{\;  1.69}$&
 $ -6.14_{ -6.17}^{ -6.09}$&
 $  1.77_{\;  1.65}^{\;  1.92}$&
 $  3.18_{\;  2.18}^{\;  6.32}$&
 $ -8.02_{ -8.28}^{ -7.49}$&
 $  0.61_{\;  0.34}^{\;  0.83}$
 \\
 050724&
 $  0.03$&
 $  2.35_{\;  2.30}^{\;  2.39}$&
 $ -6.44_{ -6.54}^{ -6.37}$&
 $  3.32_{\;  3.09}^{\;  3.62}$&
 $  5.02_{\;  4.50}^{\;  6.00}$&
 $ -6.74_{ -7.02}^{ -6.41}$&
 $  0.99_{\;  0.26}^{\;  2.03}$
 \\
 050726&
 $  0.17$&
 $  1.03_{\;  0.92}^{\;  1.12}$&
 $ -6.95_{ -6.99}^{ -6.92}$&
 $  1.15_{\;  1.13}^{\;  1.17}$&
 \\
 050730&
 $  2.03$&
 $  2.09_{\;  2.01}^{\;  2.18}$&
 $ -6.59_{ -6.61}^{ -6.57}$&
 $  1.41_{\;  1.31}^{\;  1.60}$&
 $  4.13_{\;  4.07}^{\;  4.21}$&
 $ -6.04_{ -6.15}^{ -5.95}$&
 $  2.74_{\;  2.57}^{\;  2.98}$
 \\
 050801&
 $  0.09$&
 $  1.51_{\;  1.15}^{\;  1.79}$&
 $ -7.50_{ -9.32}^{ -7.06}$&
 $  3.67_{\;  2.51}^{\; 11.55}$&
 $  2.57_{\;  2.27}^{\;  2.80}$&
 $ -7.76_{ -7.84}^{ -7.68}$&
 $  1.13_{\;  1.05}^{\;  1.20}$
 \\
 050802&
 $  1.07$&
 $  1.19_{\;  1.02}^{\;  1.41}$&
 $ -6.41_{ -6.49}^{ -6.33}$&
 $  1.63_{\;  1.50}^{\;  1.83}$&
 $  3.96_{\;  3.86}^{\;  4.05}$&
 $ -6.87_{ -6.91}^{ -6.84}$&
 $  1.66_{\;  1.54}^{\;  1.79}$
 \\
 050803&
 $ 25.50$&
 $  2.25_{\;  2.16}^{\;  2.38}$&
 $ -7.09_{ -7.62}^{ -6.87}$&
 $  4.21_{\;  3.53}^{\;  5.84}$&
 $  2.89_{ -1.50}^{\;  3.68}$&
 $ -7.24_{-10.25}^{ -6.96}$&
 $  0.66_{\;  0.60}^{\;  0.73}$
 \\
 050813&
 $  0.02$&
 $ -0.38_{ -0.67}^{ -0.01}$&
 $ -7.72_{ -7.88}^{ -7.52}$&
 $  1.63_{\;  1.48}^{\;  1.82}$&
 \\
 050814&
 $  5.16$&
 $  2.27_{\;  2.23}^{\;  2.32}$&
 $ -6.61_{ -6.68}^{ -6.55}$&
 $  2.96_{\;  2.76}^{\;  3.20}$&
 $  3.93_{\;  3.26}^{\;  4.49}$&
 $ -7.66_{ -7.99}^{ -7.50}$&
 $  0.78_{\;  0.44}^{\;  1.13}$
 \\
 050819&
 $  5.33$&
 $  1.86_{\;  1.74}^{\;  1.97}$&
 $ -6.61_{ -6.75}^{ -6.48}$&
 $  3.46_{\;  3.09}^{\;  3.89}$&
 $  4.81_{\;  3.81}^{\;  5.60}$&
 $ -7.96_{ -8.67}^{ -6.94}$&
 $  1.23_{\;  0.00}^{\; 11.92}$
 \\
 050820A&
 $  0.02$&
 $  1.45_{\;  1.22}^{\;  1.81}$&
 $ -6.93_{ -6.99}^{ -6.88}$&
 $  1.86_{\;  1.52}^{\;  2.70}$&
 $  3.96_{\;  3.84}^{\;  4.04}$&
 $ -6.36_{ -6.37}^{ -6.32}$&
 $  1.17_{\;  1.14}^{\;  1.21}$
 \\
 050822&
 $  0.13$&
 $  1.31_{\;  1.11}^{\;  1.59}$&
 $ -6.48_{ -6.57}^{ -6.37}$&
 $  1.63_{\;  1.48}^{\;  1.90}$&
 $  4.41_{\;  4.23}^{\;  4.57}$&
 $ -7.05_{ -7.12}^{ -7.00}$&
 $  0.93_{\;  0.87}^{\;  1.00}$
 \\
 050824&
 $ 52.69$&
 $  1.42_{ -1.42}^{\;  1.68}$&
 $ -6.40_{ -6.92}^{ -6.13}$&
 $  2.08_{\;  1.75}^{\;  2.71}$&
 $  4.82_{\;  4.28}^{\;  6.66}$&
 $ -7.24_{ -7.51}^{ -7.19}$&
 $  0.77_{\;  0.59}^{\;  1.25}$
 \\
 050826&
 $  1.05$&
 $  1.04_{\;  0.88}^{\;  1.20}$&
 $ -7.17_{ -7.25}^{ -7.08}$&
 $  1.12_{\;  1.09}^{\;  1.17}$&
 \\
 050904$^{\dagger}$&
 $  8.60$&
 $  2.61_{\;  2.50}^{\;  3.27}$&
 $ -6.59_{ -7.38}^{ -6.51}$&
 $  2.19_{\;  1.88}^{\;  3.53}$&
 $  4.02_{\;  3.18}^{\;  4.52}$&
 $ -7.20_{ -7.48}^{ -6.86}$&
 $  1.66_{\;  1.26}^{\;  2.17}$
 \\
 050908&
 $  1.31$&
 $  1.23_{\;  1.03}^{\;  1.50}$&
 $ -6.03_{ -6.10}^{ -5.96}$&
 $  2.39_{\;  2.10}^{\;  2.94}$&
 $  3.31_{\;  2.03}^{\;  4.17}$&
 $ -7.77_{ -8.09}^{ -7.57}$&
 $  1.07_{\;  0.78}^{\;  1.40}$
 \\
 050915A&
 $  2.58$&
 $  2.07_{\;  1.87}^{\;  2.58}$&
 $ -7.70_{-11.96}^{ -7.35}$&
 $  5.54_{\;  5.08}^{\;  6.40}$&
 $  3.09_{\;  2.33}^{\;  3.54}$&
 $ -7.81_{ -8.21}^{ -7.63}$&
 $  0.99_{\;  0.80}^{\;  1.22}$
 \\
 050915B&
 $  2.67$&
 $  2.18_{\;  2.14}^{\;  2.21}$&
 $ -6.50_{ -6.61}^{ -6.40}$&
 $  5.06_{\;  4.76}^{\;  5.37}$&
 $  2.76_{\;  1.76}^{\;  3.35}$&
 $ -8.26_{ -9.31}^{ -8.09}$&
 $  0.70_{\;  0.63}^{\;  0.76}$
 \\
 050916&
 $  8.22$&
 $  2.00_{\;  1.00}^{\;  3.00}$&
 $ -6.85_{ -6.95}^{ -6.76}$&
 $  3.32_{\;  3.04}^{\;  3.67}$&
 $  3.50_{\;  2.50}^{\;  4.22}$&
 $ -7.87_{ -8.40}^{ -7.66}$&
 $  0.74_{\;  0.39}^{\;  1.11}$
 \\
 050922B&
 $ 31.45$&
 $  2.31_{\;  2.24}^{\;  2.37}$&
 $ -5.71_{ -5.80}^{ -5.64}$&
 $  3.09_{\;  2.87}^{\;  3.33}$&
 $  5.32_{\;  4.97}^{\;  5.59}$&
 $ -6.85_{ -7.00}^{ -6.76}$&
 $  1.24_{\;  0.87}^{\;  1.70}$
 \\
 050922C&
 $  0.08$&
 $  0.38_{\;  0.17}^{\;  0.67}$&
 $ -6.84_{ -7.31}^{ -6.77}$&
 $  2.33_{\;  2.22}^{\;  4.09}$&
 $  2.58_{\;  2.48}^{\;  2.69}$&
 $ -7.03_{ -7.06}^{ -6.99}$&
 $  1.26_{\;  1.22}^{\;  1.30}$
 \\
 051001&
 $  0.81$&
 $  2.71_{\;  2.60}^{\;  2.94}$&
 $ -7.12_{ -7.91}^{ -7.07}$&
 $  4.38_{\;  3.58}^{\;  6.15}$&
 $  3.15_{\;  2.15}^{\;  5.24}$&
 $ -8.08_{ -8.45}^{ -7.75}$&
 $  0.84_{\;  0.56}^{\;  1.12}$
 \\
 051006&
 $  2.70$&
 $  1.34_{\;  1.25}^{\;  1.43}$&
 $ -6.71_{ -6.75}^{ -6.67}$&
 $  1.62_{\;  1.54}^{\;  1.70}$&
 \\
 051016A&
 $  0.45$&
 $  1.62_{\;  1.34}^{\;  2.00}$&
 $ -6.51_{ -6.84}^{ -6.42}$&
 $  2.93_{\;  2.45}^{\;  4.50}$&
 $  3.47_{\;  2.47}^{\;  4.89}$&
 $ -8.16_{-10.50}^{ -7.65}$&
 $  0.85_{\;  0.67}^{\;  1.04}$
 \\
 051016B&
 $  0.01$&
 $  0.75_{\;  0.52}^{\;  1.13}$&
 $ -6.21_{ -6.34}^{ -6.06}$&
 $  1.90_{\;  1.60}^{\;  2.48}$&
 $  3.51_{\;  2.70}^{\;  4.15}$&
 $ -6.52_{ -6.95}^{ -6.36}$&
 $  0.74_{\;  0.43}^{\;  0.93}$
 \\
 051021B&
 $  0.00$&
 $  1.35_{\;  1.25}^{\;  1.45}$&
 $ -7.13_{ -7.17}^{ -7.09}$&
 $  1.51_{\;  1.42}^{\;  1.61}$&
 \\
 051109A&
 $  0.52$&
 $  1.13_{\;  0.50}^{\;  1.45}$&
 $ -6.53_{ -6.84}^{ -6.44}$&
 $  1.52_{\;  1.34}^{\;  1.97}$&
 $  3.93_{\;  3.70}^{\;  4.08}$&
 $ -6.67_{ -6.83}^{ -6.56}$&
 $  1.25_{\;  1.18}^{\;  1.32}$
 \\
 051109B&
 $  0.71$&
 $  1.75_{\;  1.08}^{\;  2.61}$&
 $ -7.40_{ -8.56}^{ -7.23}$&
 $  3.92_{\;  2.71}^{\;  7.97}$&
 $  3.67_{\;  3.27}^{\;  3.94}$&
 $ -8.00_{ -8.20}^{ -7.89}$&
 $  1.09_{\;  0.84}^{\;  1.35}$
 \\
 051111&
 $  0.11$&
 $  1.28_{\;  1.24}^{\;  1.33}$&
 $ -5.79_{ -5.82}^{ -5.76}$&
 $  1.56_{\;  1.54}^{\;  1.58}$&
 \\
 051117A&
 $  4.95$&
 $  2.04_{\;  2.01}^{\;  2.08}$&
 $ -6.02_{ -6.04}^{ -5.98}$&
 $  1.82_{\;  1.73}^{\;  1.94}$&
 $  5.42_{\;  5.06}^{\;  5.91}$&
 $ -7.37_{ -7.50}^{ -7.28}$&
 $  1.10_{\;  0.82}^{\;  1.89}$
 \\
 051117B&
 $  1.41$&
 $  1.24_{\;  1.01}^{\;  1.44}$&
 $ -7.43_{ -7.54}^{ -7.33}$&
 $  1.60_{\;  1.46}^{\;  1.76}$&
 \\
 051210&
 $  1.24$&
 $  2.00_{\;  1.00}^{\;  3.00}$&
 $ -7.38_{ -7.42}^{ -7.33}$&
 $  2.07_{\;  1.91}^{\;  2.23}$&
 \\
 051221A&
 $  0.06$&
 $ -0.24_{ -0.29}^{ -0.19}$&
 $ -6.54_{ -6.57}^{ -4.54}$&
 $  1.49_{\;  1.47}^{\;  1.51}$&
 $  4.70_{\;  4.44}^{\;  4.94}$&
 $ -7.63_{ -7.71}^{ -7.57}$&
 $  1.32_{\;  1.08}^{\;  1.64}$
 \\
 051221B&
 $ 18.06$&
 $  1.82_{\;  1.47}^{\;  2.37}$&
 $ -7.50_{ -7.63}^{ -7.37}$&
 $  2.08_{\;  1.71}^{\;  3.39}$&
 \\
 051227$^{\dagger}$&
 $  0.02$&
 $  0.37_{\;  0.12}^{\;  0.58}$&
 $ -7.67_{ -7.79}^{ -7.56}$&
 $  1.13_{\;  1.08}^{\;  1.18}$&
 \\
 060105$^{\dagger}$$^{\ast}$&
 $  2.06$&
 $  1.08_{\;  0.98}^{\;  1.19}$&
 $ -6.15_{ -6.22}^{ -6.08}$&
 $  1.22_{\;  1.19}^{\;  1.25}$&
 $  4.30_{\;  4.24}^{\;  4.34}$&
 $ -8.93_{ -9.09}^{ -8.93}$&
 $  9.92_{\;  9.13}^{\; 19.84}$
 \\
 060108&
 $  1.05$&
 $  1.21_{\;  1.04}^{\;  1.41}$&
 $ -6.97_{ -7.03}^{ -6.90}$&
 $  2.32_{\;  2.08}^{\;  2.69}$&
 $  4.40_{\;  4.19}^{\;  4.60}$&
 $ -7.43_{ -7.50}^{ -7.36}$&
 $  1.31_{\;  1.11}^{\;  1.55}$
 \\
 060109$^{\dagger}$&
 $  0.12$&
 $  1.32_{\;  1.12}^{\;  2.00}$&
 $ -6.97_{ -7.11}^{ -6.82}$&
 $  2.05_{\;  1.71}^{\;  3.06}$&
 $  4.13_{\;  3.97}^{\;  4.29}$&
 $ -7.27_{ -7.32}^{ -7.23}$&
 $  1.57_{\;  1.35}^{\;  1.83}$
 \\
 060111A&
 $  2.81$&
 $  1.14_{\;  1.02}^{\;  1.44}$&
 $ -6.28_{-13.71}^{ -6.20}$&
 $  1.69_{\;  1.52}^{\;  2.33}$&
 $  3.33_{\;  2.33}^{\;  5.75}$&
 $ -7.83_{-10.91}^{ -6.65}$&
 $  0.77_{\;  0.51}^{\;  0.90}$
 \\
 060111B&
 $  0.17$&
 $  1.96_{\;  1.81}^{\;  2.07}$&
 $ -7.12_{ -7.32}^{ -6.93}$&
 $  3.31_{\;  2.61}^{\;  4.00}$&
 $  3.49_{\;  3.26}^{\;  3.67}$&
 $ -7.48_{ -7.56}^{ -7.41}$&
 $  1.37_{\;  1.19}^{\;  1.55}$
 \\
 060115&
 $  0.02$&
 $  1.64_{\;  1.42}^{\;  1.83}$&
 $ -6.79_{ -6.87}^{ -6.72}$&
 $  1.84_{\;  1.64}^{\;  2.08}$&
 $  3.86_{\;  2.86}^{\;  5.47}$&
 $ -7.52_{ -8.43}^{ -7.30}$&
 $  1.02_{\;  0.86}^{\;  1.22}$
 \\
 060116&
 $ 10.51$&
 $  2.25_{\;  2.10}^{\;  2.34}$&
 $ -7.60_{-10.32}^{ -7.48}$&
 $  5.89_{\;  4.87}^{\; 15.23}$&
 $  2.68_{\;  2.08}^{\;  3.09}$&
 $ -7.43_{ -7.52}^{ -7.32}$&
 $  1.11_{\;  1.01}^{\;  1.21}$
 \\
 060124&
 $  0.64$&
 $  1.34_{\;  1.18}^{\;  1.53}$&
 $ -6.73_{ -6.79}^{ -6.67}$&
 $  2.05_{\;  1.82}^{\;  2.37}$&
 $  4.60_{\;  4.47}^{\;  4.69}$&
 $ -6.22_{ -6.23}^{ -6.20}$&
 $  1.45_{\;  1.37}^{\;  1.53}$
 \\
 060202$^{\dagger}$&
 $  0.03$&
 $  2.14_{\;  1.81}^{\;  2.39}$&
 $ -6.22_{ -6.37}^{ -6.10}$&
 $  1.59_{\;  1.39}^{\;  1.84}$&
 $  4.99_{\;  4.70}^{\;  5.34}$&
 $ -6.92_{ -7.02}^{ -6.82}$&
 $  0.94_{\;  0.76}^{\;  1.18}$
 \\
 060204B&
 $  2.72$&
 $  1.24_{\;  1.06}^{\;  1.53}$&
 $ -6.69_{ -6.78}^{ -6.59}$&
 $  1.36_{\;  1.23}^{\;  1.64}$&
 $  4.13_{\;  3.95}^{\;  4.33}$&
 $ -7.16_{ -7.30}^{ -7.07}$&
 $  1.48_{\;  1.22}^{\;  1.82}$
 \\
 060206&
 $  1.41$&
 $  0.53_{\;  0.46}^{\;  0.61}$&
 $ -6.93_{ -6.97}^{ -6.89}$&
 $  1.47_{\;  1.41}^{\;  1.53}$&
 $  3.86_{\;  3.68}^{\;  4.00}$&
 $ -6.55_{ -6.61}^{ -6.50}$&
 $  1.24_{\;  1.18}^{\;  1.29}$
 \\
 060210$^{\dagger}$$^{\ast}$&
 $  0.01$&
 $  0.43_{\;  0.27}^{\;  0.62}$&
 $ -6.87_{ -6.97}^{ -6.76}$&
 $  1.00_{\;  0.97}^{\;  1.07}$&
 $  4.46_{\;  4.17}^{\;  4.79}$&
 $ -6.85_{ -7.16}^{ -6.75}$&
 $  1.76_{\;  1.51}^{\;  2.86}$
 \\
 060211A&
 $ 28.76$&
 $  2.39_{\;  2.34}^{\;  2.44}$&
 $ -6.62_{ -6.76}^{ -6.52}$&
 $  3.65_{\;  3.38}^{\;  4.01}$&
 $  2.99_{\;  1.99}^{\;  4.48}$&
 $ -8.11_{-10.29}^{ -7.56}$&
 $  0.88_{\;  0.68}^{\;  1.04}$
 \\
 060211B&
 $  0.01$&
 $  1.38_{\;  1.17}^{\;  1.59}$&
 $ -7.23_{ -7.31}^{ -7.16}$&
 $  1.91_{\;  1.68}^{\;  2.22}$&
 $  4.00_{\;  3.00}^{\;  5.00}$&
 $ -8.39_{ -8.62}^{ -8.18}$&
 $  0.46_{\;  0.00}^{\;  1.64}$
 \\
 060218&
 $  0.81$&
 $  3.05_{\;  2.54}^{\;  3.26}$&
 $ -6.46_{ -7.40}^{ -6.37}$&
 $  1.89_{\;  1.82}^{\;  6.72}$&
 $  5.01_{\;  4.53}^{\;  5.64}$&
 $ -7.45_{ -7.94}^{ -7.03}$&
 $  1.29_{\;  0.77}^{\;  2.35}$
 \\
 060219&
 $  3.40$&
 $  1.14_{\;  0.78}^{\;  1.65}$&
 $ -6.00_{ -6.13}^{ -5.88}$&
 $  2.65_{\;  2.29}^{\;  3.78}$&
 $  4.59_{\;  4.04}^{\;  4.95}$&
 $ -7.38_{ -7.52}^{ -7.27}$&
 $  1.41_{\;  0.93}^{\;  1.96}$
 \\
 060223A&
 $  0.09$&
 $  1.44_{\;  1.22}^{\;  2.09}$&
 $ -7.30_{-10.20}^{ -6.90}$&
 $  3.82_{\;  3.07}^{\; 13.13}$&
 $  2.73_{\;  2.37}^{\;  3.05}$&
 $ -8.19_{ -8.29}^{ -8.10}$&
 $  1.30_{\;  1.12}^{\;  1.50}$
 \\
 060306$^{\dagger}$&
 $  0.43$&
 $ -0.50_{ -1.50}^{\;  0.09}$&
 $ -6.29_{ -6.99}^{ -4.29}$&
 $  1.46_{\;  1.40}^{\;  1.54}$&
 $  3.90_{\;  3.71}^{\;  4.10}$&
 $ -6.91_{ -6.98}^{ -6.86}$&
 $  0.99_{\;  0.88}^{\;  1.11}$
 \\
 060312&
 $  1.18$&
 $  1.62_{\;  1.48}^{\;  1.83}$&
 $ -6.73_{ -7.24}^{ -6.67}$&
 $  3.18_{\;  2.75}^{\;  4.65}$&
 $  3.15_{\;  2.68}^{\;  3.45}$&
 $ -7.66_{ -7.84}^{ -7.53}$&
 $  1.02_{\;  0.88}^{\;  1.20}$
 \\
 060313&
 $  0.20$&
 $ -0.13_{ -0.34}^{\;  0.14}$&
 $ -7.30_{ -7.44}^{ -7.11}$&
 $  0.79_{\;  0.74}^{\;  0.84}$&
 \\
 060319&
 $  4.25$&
 $  0.93_{\;  0.62}^{\;  1.32}$&
 $ -6.67_{ -6.81}^{ -6.52}$&
 $  1.64_{\;  1.44}^{\;  2.08}$&
 $  3.18_{\;  2.59}^{\;  3.59}$&
 $ -7.65_{ -7.99}^{ -7.41}$&
 $  0.86_{\;  0.53}^{\;  1.12}$
 \\
 060323&
 $  2.55$&
 $  1.60_{\;  1.30}^{\;  2.47}$&
 $ -7.33_{ -8.18}^{ -7.23}$&
 $  2.77_{\;  2.69}^{\;  3.00}$&
 $  3.29_{\;  2.89}^{\;  3.59}$&
 $ -7.76_{ -7.90}^{ -7.68}$&
 $  1.38_{\;  1.14}^{\;  1.66}$
 \\
 060403&
 $  3.88$&
 $  1.55_{\;  1.48}^{\;  1.62}$&
 $ -6.95_{ -6.96}^{ -6.94}$&
 $  1.40_{\;  1.33}^{\;  1.47}$&
 \\
 060413&
 $ 24.50$&
 $  2.24_{ -0.74}^{\;  2.28}$&
 $ -6.10_{ -6.23}^{ -6.02}$&
 $  3.16_{\;  2.87}^{\;  3.53}$&
 $  4.58_{\;  4.44}^{\;  4.70}$&
 $ -6.38_{ -6.43}^{ -6.33}$&
 $  1.96_{\;  1.69}^{\;  2.26}$
 \\
 060418&
 $  2.39$&
 $  1.75_{\;  1.70}^{\;  1.79}$&
 $ -5.73_{ -5.77}^{ -5.72}$&
 $  2.35_{\;  2.23}^{\;  2.49}$&
 $  3.44_{\;  3.26}^{\;  3.58}$&
 $ -6.87_{ -6.95}^{ -6.82}$&
 $  1.48_{\;  1.33}^{\;  1.62}$
 \\
 060421&
 $  0.88$&
 $  1.15_{\;  0.99}^{\;  1.84}$&
 $ -6.99_{ -8.30}^{ -6.92}$&
 $  2.36_{\;  2.00}^{\;  7.88}$&
 $  3.04_{\;  2.54}^{\;  3.43}$&
 $ -7.31_{ -7.49}^{ -7.17}$&
 $  1.25_{\;  1.03}^{\;  1.55}$
 \\
 060427&
 $  0.20$&
 $  2.31_{\;  2.30}^{\;  2.39}$&
 $ -7.04_{ -7.18}^{ -6.82}$&
 $  5.15_{\;  4.55}^{\;  7.05}$&
 $  2.69_{\;  1.69}^{\;  3.82}$&
 $ -7.44_{ -7.62}^{ -7.03}$&
 $  1.41_{\;  1.29}^{\;  1.53}$
 \\
 060428A&
 $  3.04$&
 $  1.12_{\;  0.98}^{\;  1.28}$&
 $ -5.61_{ -5.74}^{ -5.53}$&
 $  2.59_{\;  2.27}^{\;  3.17}$&
 $  3.38_{\;  2.99}^{\;  3.61}$&
 $ -6.49_{ -6.73}^{ -6.38}$&
 $  0.65_{\;  0.55}^{\;  0.72}$
 \\
 060428B&
 $  5.86$&
 $  2.26_{\;  2.22}^{\;  2.29}$&
 $ -6.04_{ -6.14}^{ -5.94}$&
 $  4.57_{\;  4.33}^{\;  4.82}$&
 $  3.50_{\;  3.06}^{\;  3.86}$&
 $ -7.88_{ -8.03}^{ -7.78}$&
 $  1.02_{\;  0.88}^{\;  1.16}$
 \\
 060502A&
 $  1.02$&
 $  1.63_{\;  1.52}^{\;  1.78}$&
 $ -6.56_{ -6.78}^{ -6.53}$&
 $  2.68_{\;  2.43}^{\;  3.68}$&
 $  4.25_{\;  3.36}^{\;  4.49}$&
 $ -7.10_{ -7.46}^{ -7.01}$&
 $  1.00_{\;  0.76}^{\;  1.26}$
 \\
 060502B&
 $  0.00$&
 $ -1.16_{ -1.44}^{ -0.78}$&
 $ -8.48_{ -8.62}^{ -8.27}$&
 $  1.24_{\;  1.16}^{\;  1.34}$&
 \\
 060510A&
 $  0.01$&
 $  1.40_{\;  1.29}^{\;  1.54}$&
 $ -5.21_{ -5.47}^{ -5.07}$&
 $  4.34_{\;  3.84}^{\;  5.21}$&
 $  4.11_{\;  4.04}^{\;  4.18}$&
 $ -5.49_{ -5.52}^{ -5.45}$&
 $  1.51_{\;  1.43}^{\;  1.60}$
 \\
 060510B$^{\dagger}$&
 $117.20$&
 $  2.60_{\;  2.46}^{\;  2.79}$&
 $ -7.20_{ -8.53}^{ -7.16}$&
 $  4.08_{\;  3.16}^{\; 11.33}$&
 $  4.55_{\;  4.01}^{\;  5.15}$&
 $ -7.90_{ -8.15}^{ -7.79}$&
 $  0.96_{\;  0.72}^{\;  1.40}$
 \\
 060512&
 $  0.35$&
 $  1.18_{\;  0.90}^{\;  2.40}$&
 $ -5.69_{ -5.92}^{ -5.58}$&
 $  2.99_{\;  2.54}^{\;  9.41}$&
 $  3.85_{\;  2.85}^{\;  4.19}$&
 $ -8.05_{ -8.14}^{ -7.87}$&
 $  1.17_{\;  1.01}^{\;  1.38}$
 \\
 060522&
 $  7.00$&
 $  2.36_{\;  2.23}^{\;  2.44}$&
 $ -7.80_{ -9.95}^{ -7.62}$&
 $  4.97_{\;  3.86}^{\;  9.16}$&
 $  2.86_{\;  2.61}^{\;  3.32}$&
 $ -7.81_{ -7.90}^{ -7.73}$&
 $  1.04_{\;  0.93}^{\;  1.20}$
 \\
 060526&
 $  0.62$&
 $  1.20_{\;  1.03}^{\;  1.39}$&
 $ -6.82_{ -6.89}^{ -6.76}$&
 $  1.87_{\;  1.68}^{\;  2.13}$&
 $  3.84_{\;  3.51}^{\;  4.31}$&
 $ -7.39_{ -7.48}^{ -7.25}$&
 $  1.06_{\;  0.89}^{\;  1.37}$
 \\
 060604&
 $  0.02$&
 $  1.49_{\;  1.10}^{\;  1.92}$&
 $ -6.49_{ -6.66}^{ -6.35}$&
 $  1.56_{\;  1.33}^{\;  2.10}$&
 $  4.55_{\;  4.26}^{\;  4.83}$&
 $ -7.20_{ -7.34}^{ -7.12}$&
 $  1.18_{\;  0.99}^{\;  1.38}$
 \\
 060605&
 $  2.02$&
 $  1.44_{\;  1.33}^{\;  1.82}$&
 $ -7.33_{ -7.43}^{ -7.26}$&
 $  1.83_{\;  1.58}^{\;  2.58}$&
 $  4.16_{\;  4.03}^{\;  4.29}$&
 $ -7.15_{ -7.24}^{ -7.08}$&
 $  2.04_{\;  1.76}^{\;  2.39}$
 \\
 060607A$^{\dagger}$$^{\ast}$&
 $  1.01$&
 $  1.09_{\;  0.98}^{\;  1.23}$&
 $ -6.45_{ -6.51}^{ -6.38}$&
 $  1.33_{\;  1.27}^{\;  1.40}$&
 $  4.75_{\;  4.73}^{\;  4.78}$&
 $ -8.38_{ -8.80}^{ -6.82}$&
 $  8.48_{\;  8.14}^{\; 16.97}$
 \\
 060614$^{\dagger}$&
 $  0.29$&
 $  0.70_{\;  0.58}^{\;  0.85}$&
 $ -6.21_{ -6.27}^{ -6.16}$&
 $  1.61_{\;  1.47}^{\;  1.97}$&
 $  5.00_{\;  4.91}^{\;  5.08}$&
 $ -6.60_{ -6.65}^{ -6.56}$&
 $  1.98_{\;  1.78}^{\;  2.22}$
 \\
 060707&
 $  1.03$&
 $  1.45_{\;  1.27}^{\;  1.55}$&
 $ -6.63_{ -6.69}^{ -6.58}$&
 $  1.93_{\;  1.71}^{\;  2.11}$&
 $  3.58_{\;  2.87}^{\;  4.05}$&
 $ -7.46_{ -7.70}^{ -7.20}$&
 $  0.83_{\;  0.75}^{\;  0.91}$
 \\
 060708&
 $  0.74$&
 $  1.82_{\;  1.77}^{\;  1.88}$&
 $ -6.31_{ -6.47}^{ -6.20}$&
 $  4.18_{\;  3.80}^{\;  4.64}$&
 $  3.41_{\;  3.16}^{\;  3.66}$&
 $ -7.13_{ -7.20}^{ -7.06}$&
 $  1.08_{\;  1.00}^{\;  1.15}$
 \\
 060712&
 $  9.79$&
 $  1.57_{\;  1.30}^{\;  1.87}$&
 $ -6.42_{ -6.53}^{ -6.30}$&
 $  2.55_{\;  2.26}^{\;  3.18}$&
 $  4.12_{\;  3.51}^{\;  4.39}$&
 $ -7.47_{ -7.60}^{ -7.39}$&
 $  1.08_{\;  0.90}^{\;  1.27}$
 \\
 060714&
 $  0.02$&
 $  1.87_{\;  1.64}^{\;  2.19}$&
 $ -6.40_{ -6.54}^{ -6.34}$&
 $  2.43_{\;  2.07}^{\;  3.40}$&
 $  3.20_{\;  2.79}^{\;  3.41}$&
 $ -6.73_{ -6.80}^{ -6.63}$&
 $  1.25_{\;  1.18}^{\;  1.34}$
 \\
 060717&
 $  0.02$&
 $  1.55_{\;  1.28}^{\;  1.89}$&
 $ -7.17_{ -7.34}^{ -7.00}$&
 $  2.96_{\;  2.33}^{\;  3.96}$&
 $  3.40_{\;  2.40}^{\;  4.61}$&
 $ -8.88_{-10.98}^{ -6.66}$&
 $  0.73_{\;  0.00}^{\;  2.14}$
 \\
 060719&
 $  0.01$&
 $  1.98_{\;  1.91}^{\;  2.07}$&
 $ -6.70_{ -7.17}^{ -6.43}$&
 $  5.93_{\;  5.11}^{\;  7.12}$&
 $  3.81_{\;  3.47}^{\;  4.07}$&
 $ -7.08_{ -7.16}^{ -7.02}$&
 $  1.10_{\;  0.98}^{\;  1.24}$
 \\
 060729&
 $  0.01$&
 $  1.55_{\;  1.33}^{\;  1.71}$&
 $ -5.99_{ -6.12}^{ -5.88}$&
 $  2.22_{\;  1.99}^{\;  2.44}$&
 $  5.11_{\;  5.07}^{\;  5.15}$&
 $ -6.09_{ -6.13}^{ -6.07}$&
 $  1.32_{\;  1.27}^{\;  1.37}$
 \\
 060801&
 $  0.00$&
 $ -0.30_{ -0.52}^{ -0.03}$&
 $ -8.45_{ -8.57}^{ -8.28}$&
 $  0.95_{\;  0.91}^{\;  1.00}$&
 \\
 \enddata
 \end{deluxetable}

%% file: tab2.tex
 \begin{deluxetable}{lrrrrr}
 \tablewidth{0pt}
 \tablecaption{
 Late temporal breaks with 90\% confidence limits.
 $T_{b}$ is the break time in seconds.
 The decay index before the break was either $\alpha_{p}$ or
 $\alpha_{a}$ as shown depending on which component dominates
 at the end of the X-ray light curve.
 The final decay index after the late break is given by $\alpha_{b}$.
 $\Delta\alpha$ is the change in decay index across the break.
 For 060607A the late break is close to the decaying section of
 the 2nd component so the $\Delta\alpha$ value maybe
 misleading/incorrect.\label{tab2}}
 \tablehead{
 \colhead{GRB}&\colhead{$\alpha_{p}$}&
 \colhead{$\alpha_{a}$}&\colhead{$\log_{10}(T_{b})$}&
 \colhead{$\alpha_{b}$}&\colhead{$\Delta\alpha$}}
 \startdata
 050315&
 &$  0.72$&
 $  5.48_{\;  5.10}^{\;  5.75}$&
 $  2.00_{\;  0.86}^{\;  4.07}$&
 $  1.28_{\;  0.14}^{\;  3.35}$\\
 050814&
 &$  0.78$&
 $  4.93_{\;  4.41}^{\;  5.36}$&
 $  1.80_{\;  1.01}^{\;  3.16}$&
 $  1.02_{\;  0.23}^{\;  2.38}$\\
 051016B&
 &$  0.74$&
 $  4.57_{\;  4.00}^{\;  5.16}$&
 $  1.23_{\;  0.94}^{\;  1.57}$&
 $  0.49_{\;  0.19}^{\;  0.83}$\\
 060105&
 $  1.22$&&
 $  5.04_{\;  4.66}^{\;  5.26}$&
 $  3.56_{\;  1.45}^{\;  8.29}$&
 $  2.33_{\;  0.23}^{\;  7.07}$\\
 060313&
 $  0.79$&&
 $  3.87_{\;  3.52}^{\;  4.04}$&
 $  1.64_{\;  1.37}^{\;  1.92}$&
 $  0.85_{\;  0.58}^{\;  1.13}$\\
 060319&
 &$  0.86$&
 $  4.77_{\;  4.06}^{\;  5.39}$&
 $  1.21_{\;  1.00}^{\;  1.44}$&
 $  0.34_{\;  0.14}^{\;  0.58}$\\
 060428A&
 &$  0.65$&
 $  5.14_{\;  4.84}^{\;  5.34}$&
 $  1.41_{\;  1.10}^{\;  1.79}$&
 $  0.76_{\;  0.45}^{\;  1.14}$\\
 060607A&
 $  1.33$&&
 $  4.81_{\;  4.06}^{\; 63.51}$&
 $  4.47_{\;  0.00}^{\;  8.94}$&
 $  3.14_{\; -1.33}^{\;  7.61}$\\
 \enddata
 \end{deluxetable}

%% file: tab3.tex
 \begin{deluxetable}{lrrrrrr}
 \tablewidth{0pt}
 \tablecaption{
 Spectral indices;
 $\beta_{p}$ prompt phase,
 $\beta_{pd}$ initial decay,
 $\beta_{a}$ plateau
 and $\beta_{ad}$ final decay after the plateau.
 Limits are 90\% confidence.
 Column D is the type of decay fit from Table \ref{tab1};
 1 - components p and a required with component a
 dominant at end,
 2 - components p and a required with component p dominant at end,
 3 - only component p required.
 Column z is the measured redshift taken from
 http://swift.gsfc.nasa.gov/docs/swift/archive/grb\_table/.
 References for all the redshifts are provided on this WWW data
 table.\label{tab3}}
 \tablehead{
 \colhead{GRB}&\colhead{D}&
 \colhead{$\beta_{p}$}&\colhead{$\beta_{pd}$}&
 \colhead{$\beta_{a}$}&\colhead{$\beta_{ad}$}&
 \colhead{z}}
 \startdata
 050126&
 1&
 $ 0.41\pm 0.15$&
 $ 1.59\pm 0.38$&
 &
 $ 0.72\pm 0.62$&
 1.290\\
 050128&
 1&
 $-0.53\pm 0.36$&
 $ 0.85\pm 0.12$&
 $ 0.96\pm 0.11$&
 $ 1.13\pm 0.12$&
 \\
 050219A&
 1&
 $-1.03\pm 0.29$&
 $ 1.02\pm 0.20$&
 &
 $ 0.89\pm 0.24$&
 \\
 050315&
 1&
 $ 1.15\pm 0.09$&
 $ 1.49\pm 0.19$&
 $ 0.89\pm 0.07$&
 $ 1.29\pm 0.31$&
 1.949\\
 050318&
 1&
 $ 1.01\pm 0.10$&
 $ 0.93\pm 0.30$&
 &
 &
 1.440\\
 050319&
 1&
 $ 1.10\pm 0.20$&
 $ 2.02\pm 0.47$&
 $ 0.85\pm 0.04$&
 $ 1.36\pm 0.54$&
 3.240\\
 050401&
 2&
 $ 0.52\pm 0.07$&
 $ 0.98\pm 0.05$&
 $ 1.00\pm 0.06$&
 $ 0.89\pm 0.20$&
 2.900\\
 050406&
 1&
 $ 1.64\pm 0.47$&
 $ 1.37\pm 0.25$&
 &
 &
 2.440\\
 050412&
 2&
 $-0.26\pm 0.18$&
 $ 0.26\pm 0.32$&
 &
 &
 \\
 050416A&
 1&
 $ 2.20\pm 0.25$&
 $ 0.80\pm 0.29$&
 $ 0.99\pm 0.17$&
 $ 0.93\pm 0.08$&
 0.654\\
 050421&
 2&
 $ 0.64\pm 0.46$&
 $ 0.27\pm 0.36$&
 $-0.11\pm 0.50$&
 $ 0.69\pm 0.18$&
 \\
 050422&
 1&
 $ 0.54\pm 0.21$&
 $ 2.23\pm 0.60$&
 &
 $ 1.08\pm 0.69$&
 \\
 050502B&
 1&
 $ 0.64\pm 0.15$&
 $ 0.81\pm 0.28$&
 &
 &
 \\
 050505&
 1&
 $ 0.55\pm 0.12$&
 $ 0.80\pm 0.07$&
 $ 1.09\pm 0.07$&
 $ 1.24\pm 0.19$&
 4.270\\
 050509B&
 3&
 $ 0.47\pm 0.25$&
 $ 0.47\pm 0.25$&
 &
 &
 0.225\\
 050525A&
 1&
 $-0.17\pm 0.12$&
 $ 1.07\pm 0.02$&
 &
 $ 1.13\pm 0.22$&
 0.606\\
 050603&
 1&
 $ 0.11\pm 0.06$&
 $ 0.75\pm 0.26$&
 $ 0.91\pm 0.16$&
 $ 0.70\pm 0.10$&
 2.821\\
 050607&
 1&
 $ 0.97\pm 0.17$&
 $ 0.77\pm 0.48$&
 $ 0.74\pm 0.18$&
 $ 1.28\pm 0.27$&
 \\
 050712&
 1&
 $ 0.50\pm 0.19$&
 $ 0.91\pm 0.06$&
 $ 0.90\pm 0.12$&
 $ 0.80\pm 0.26$&
 \\
 050713A&
 1&
 $ 0.55\pm 0.07$&
 $ 1.30\pm 0.07$&
 $ 1.32\pm 0.17$&
 $ 0.86\pm 0.22$&
 \\
 050713B&
 1&
 $ 0.53\pm 0.15$&
 $ 0.70\pm 0.11$&
 $ 1.05\pm 0.12$&
 $ 0.90\pm 0.07$&
 \\
 050714B&
 1&
 $ 1.70\pm 0.41$&
 $ 1.70\pm 0.41$&
 &
 $ 1.44\pm 0.45$&
 \\
 050716&
 1&
 $-0.17\pm 0.28$&
 $ 0.33\pm 0.03$&
 $ 0.84\pm 0.20$&
 $ 0.93\pm 0.14$&
 \\
 050717&
 3&
 $ 0.36\pm 0.05$&
 $ 0.63\pm 0.11$&
 &
 $ 0.35\pm 0.20$&
 \\
 050721&
 1&
 $ 0.78\pm 0.12$&
 $ 0.74\pm 0.15$&
 &
 $ 0.40\pm 0.08$&
 \\
 050724&
 1&
 $ 1.17\pm 0.26$&
 $ 0.95\pm 0.07$&
 &
 &
 0.257\\
 050726&
 3&
 $ 0.01\pm 0.17$&
 $ 0.94\pm 0.07$&
 &
 $ 0.93\pm 0.07$&
 \\
 050730&
 1&
 $ 0.52\pm 0.11$&
 $ 0.33\pm 0.08$&
 &
 $ 0.62\pm 0.05$&
 3.970\\
 050801&
 1&
 $ 1.03\pm 0.24$&
 $ 0.72\pm 0.54$&
 &
 $ 0.84\pm 0.14$&
 \\
 050802&
 1&
 $ 0.66\pm 0.15$&
 $ 0.91\pm 0.19$&
 $ 0.89\pm 0.11$&
 $ 0.81\pm 0.09$&
 1.710\\
 050803&
 1&
 $ 0.47\pm 0.11$&
 $ 0.71\pm 0.16$&
 &
 $ 0.92\pm 0.12$&
 0.422\\
 050813&
 3&
 $ 0.37\pm 0.37$&
 $ 1.42\pm 0.86$&
 &
 &
 1.800\\
 050814&
 1&
 $ 0.98\pm 0.19$&
 $ 1.08\pm 0.08$&
 &
 $ 0.71\pm 0.10$&
 5.300\\
 050819&
 1&
 $ 1.56\pm 0.21$&
 $ 1.18\pm 0.23$&
 &
 $ 0.55\pm 0.47$&
 \\
 050820A&
 1&
 $ 0.24\pm 0.07$&
 $ 0.87\pm 0.09$&
 $ 1.28\pm 0.05$&
 $ 0.74\pm 0.75$&
 2.612\\
 050822&
 1&
 $ 1.53\pm 0.09$&
 $ 1.60\pm 0.06$&
 $ 1.24\pm 0.14$&
 $ 1.13\pm 0.10$&
 \\
 050824&
 1&
 $ 1.90\pm 0.42$&
 $ 0.91\pm 0.15$&
 $ 0.84\pm 0.13$&
 &
 0.830\\
 050826&
 3&
 $ 0.10\pm 0.28$&
 $ 1.27\pm 0.47$&
 &
 $ 1.75\pm 0.38$&
 \\
 050904&
 1&
 $ 0.38\pm 0.04$&
 $ 0.44\pm 0.04$&
 $ 0.61\pm 0.04$&
 $ 1.00\pm 0.14$&
 6.290\\
 050908&
 1&
 $ 0.91\pm 0.11$&
 $ 2.35\pm 0.27$&
 $ 0.80\pm 0.27$&
 &
 3.350\\
 050915A&
 1&
 $ 0.37\pm 0.11$&
 $ 1.12\pm 0.34$&
 $ 1.06\pm 0.23$&
 &
 \\
 050915B&
 1&
 $ 0.89\pm 0.06$&
 $ 1.45\pm 0.10$&
 $ 0.93\pm 0.18$&
 $ 0.71\pm 0.95$&
 \\
 050916&
 1&
 $ 0.83\pm 0.32$&
 $ 0.77\pm 0.84$&
 &
 &
 \\
 050922B&
 1&
 $ 1.11\pm 0.16$&
 $ 1.64\pm 0.08$&
 $ 1.34\pm 0.55$&
 $ 1.33\pm 0.25$&
 \\
 050922C&
 1&
 $ 0.34\pm 0.03$&
 $ 1.10\pm 0.09$&
 $ 0.91\pm 0.38$&
 $ 1.32\pm 0.16$&
 2.198\\
 051001&
 1&
 $ 1.19\pm 0.10$&
 $ 0.49\pm 0.06$&
 $ 1.45\pm 0.91$&
 $ 1.43\pm 0.79$&
 \\
 051006&
 3&
 $ 0.68\pm 0.16$&
 $ 0.54\pm 0.51$&
 &
 &
 \\
 051016A&
 1&
 $ 0.95\pm 0.16$&
 $ 1.16\pm 0.73$&
 &
 $ 1.07\pm 0.49$&
 \\
 051016B&
 1&
 $ 1.53\pm 0.17$&
 $ 1.89\pm 0.81$&
 $ 0.86\pm 0.21$&
 $ 1.26\pm 0.33$&
 0.936\\
 051021B&
 3&
 $ 0.57\pm 0.10$&
 $ 0.39\pm 0.46$&
 &
 &
 \\
 051109A&
 1&
 $ 0.53\pm 0.15$&
 $ 0.98\pm 0.12$&
 $ 0.93\pm 0.04$&
 $ 1.04\pm 0.12$&
 2.346\\
 051109B&
 1&
 $ 0.98\pm 0.19$&
 $ 1.00\pm 0.28$&
 $ 0.55\pm 0.30$&
 $ 0.94\pm 0.32$&
 \\
 051111&
 3&
 $ 0.38\pm 0.11$&
 $ 1.28\pm 0.24$&
 &
 $ 1.07\pm 0.36$&
 1.549\\
 051117A&
 1&
 $ 0.76\pm 0.03$&
 $ 1.06\pm 0.02$&
 $ 1.55\pm 0.21$&
 $ 1.73\pm 0.45$&
 \\
 051117B&
 3&
 $ 0.74\pm 0.25$&
 $ 0.34\pm 0.31$&
 &
 &
 \\
 051210&
 3&
 $ 0.00\pm 0.00$&
 $ 0.26\pm 0.12$&
 &
 &
 \\
 051221A&
 1&
 $ 0.34\pm 0.04$&
 $ 0.83\pm 0.11$&
 &
 $ 0.68\pm 0.19$&
 0.547\\
 051221B&
 3&
 $ 0.32\pm 0.13$&
 $ 0.35\pm 0.51$&
 &
 &
 \\
 051227&
 3&
 $ 0.57\pm 0.17$&
 $ 0.41\pm 0.22$&
 &
 $ 0.58\pm 0.40$&
 \\
 060105&
 2&
 $ 0.07\pm 0.02$&
 $ 1.11\pm 0.03$&
 &
 $ 1.07\pm 0.03$&
 \\
 060108&
 1&
 $ 0.94\pm 0.11$&
 $ 0.98\pm 0.25$&
 $ 0.68\pm 0.22$&
 &
 2.030\\
 060109&
 1&
 $ 0.99\pm 0.18$&
 $ 1.16\pm 0.17$&
 &
 $ 1.08\pm 0.15$&
 \\
 060111A&
 1&
 $ 0.67\pm 0.05$&
 $ 1.35\pm 0.04$&
 &
 $ 1.39\pm 0.12$&
 \\
 060111B&
 1&
 $ 0.08\pm 0.12$&
 $ 1.16\pm 0.21$&
 $ 1.10\pm 0.12$&
 $ 1.04\pm 0.16$&
 \\
 060115&
 1&
 $ 0.79\pm 0.08$&
 $ 0.72\pm 0.06$&
 $ 1.24\pm 0.16$&
 $ 1.28\pm 0.23$&
 3.530\\
 060116&
 1&
 $ 0.36\pm 0.13$&
 $ 1.20\pm 0.40$&
 $ 0.93\pm 0.32$&
 $ 1.53\pm 0.58$&
 6.600\\
 060124&
 1&
 $ 0.89\pm 0.13$&
 $ 0.86\pm 0.11$&
 $ 1.01\pm 0.06$&
 $ 1.28\pm 0.11$&
 2.300\\
 060202&
 1&
 $ 0.81\pm 0.09$&
 $ 1.32\pm 0.03$&
 $ 2.17\pm 0.20$&
 $ 1.97\pm 0.21$&
 \\
 060204B&
 1&
 $ 0.38\pm 0.06$&
 $ 0.92\pm 0.09$&
 $ 1.33\pm 0.21$&
 $ 1.47\pm 0.30$&
 \\
 060206&
 1&
 $ 0.76\pm 0.05$&
 $ 0.63\pm 0.33$&
 $ 0.78\pm 0.23$&
 $ 0.80\pm 0.16$&
 4.050\\
 060210&
 2&
 $ 0.69\pm 0.12$&
 $ 1.02\pm 0.04$&
 $ 1.05\pm 0.07$&
 $ 1.00\pm 0.09$&
 3.910\\
 060211A&
 1&
 $ 0.83\pm 0.08$&
 $ 0.99\pm 0.08$&
 &
 $ 0.94\pm 0.41$&
 \\
 060211B&
 1&
 $ 0.58\pm 0.15$&
 $ 0.92\pm 0.45$&
 &
 $ 1.50\pm 0.81$&
 \\
 060218&
 1&
 $ 1.37\pm 0.25$&
 $ 1.72\pm 0.19$&
 &
 $ 3.00\pm 0.18$&
 0.030\\
 060219&
 1&
 $ 1.65\pm 0.28$&
 $ 2.15\pm 1.06$&
 $ 2.04\pm 0.57$&
 $ 1.61\pm 1.00$&
 \\
 060223A&
 1&
 $ 0.77\pm 0.08$&
 $ 0.90\pm 0.23$&
 $ 1.02\pm 0.19$&
 $ 0.85\pm 0.25$&
 4.410\\
 060306&
 1&
 $ 0.83\pm 0.07$&
 $ 1.04\pm 0.23$&
 $ 1.17\pm 0.19$&
 $ 1.18\pm 0.17$&
 \\
 060312&
 1&
 $ 0.87\pm 0.05$&
 $ 1.27\pm 0.10$&
 $ 1.08\pm 0.29$&
 $ 0.88\pm 0.36$&
 \\
 060313&
 3&
 $-0.37\pm 0.05$&
 $ 0.66\pm 0.16$&
 &
 $ 1.34\pm 0.31$&
 \\
 060319&
 1&
 $ 1.29\pm 0.15$&
 $ 1.16\pm 0.18$&
 $ 1.18\pm 0.17$&
 $ 1.21\pm 0.81$&
 \\
 060323&
 1&
 $ 0.51\pm 0.11$&
 $ 0.74\pm 0.18$&
 &
 &
 \\
 060403&
 3&
 $-0.01\pm 0.06$&
 $ 0.64\pm 0.21$&
 &
 $ 1.50\pm 0.75$&
 \\
 060413&
 1&
 $ 0.70\pm 0.04$&
 $ 0.86\pm 0.10$&
 $ 1.46\pm 0.39$&
 $ 0.53\pm 0.12$&
 \\
 060418&
 1&
 $ 0.66\pm 0.03$&
 $ 1.26\pm 0.06$&
 $ 1.04\pm 0.21$&
 $ 0.81\pm 0.89$&
 1.490\\
 060421&
 1&
 $ 0.49\pm 0.05$&
 $ 0.54\pm 0.33$&
 $ 0.41\pm 0.39$&
 $ 0.33\pm 0.37$&
 \\
 060427&
 1&
 $ 0.90\pm 0.17$&
 $ 1.90\pm 0.21$&
 $ 0.54\pm 0.34$&
 $ 1.15\pm 0.17$&
 \\
 060428A&
 1&
 $ 1.04\pm 0.07$&
 $ 2.35\pm 0.25$&
 $ 1.20\pm 0.15$&
 $ 0.94\pm 0.12$&
 \\
 060428B&
 1&
 $ 1.70\pm 0.14$&
 $ 1.80\pm 0.11$&
 $ 2.35\pm 0.25$&
 $ 0.86\pm 0.07$&
 \\
 060502A&
 1&
 $ 0.45\pm 0.05$&
 $ 2.37\pm 0.34$&
 $ 1.21\pm 0.19$&
 $ 1.07\pm 0.27$&
 1.510\\
 060502B&
 3&
 $ 0.26\pm 0.25$&
 $ 0.95\pm 0.55$&
 &
 $ 0.62\pm 0.56$&
 \\
 060510A&
 1&
 $ 0.76\pm 0.08$&
 $ 2.73\pm 0.30$&
 $ 1.05\pm 0.07$&
 $ 0.97\pm 0.08$&
 \\
 060510B&
 1&
 $ 0.78\pm 0.05$&
 $ 1.27\pm 0.33$&
 $ 1.57\pm 0.35$&
 $ 1.67\pm 0.36$&
 4.900\\
 060512&
 1&
 $ 1.43\pm 0.22$&
 $ 0.88\pm 0.16$&
 &
 &
 0.443\\
 060522&
 1&
 $ 0.55\pm 0.11$&
 $ 0.95\pm 0.13$&
 $ 0.82\pm 0.22$&
 &
 5.110\\
 060526&
 1&
 $ 1.00\pm 0.15$&
 $ 0.87\pm 0.10$&
 $ 0.89\pm 0.14$&
 $ 0.13\pm 0.62$&
 3.210\\
 060604&
 1&
 $ 0.88\pm 0.30$&
 $ 2.72\pm 0.19$&
 $ 1.36\pm 0.20$&
 $ 1.25\pm 0.16$&
 2.680\\
 060605&
 1&
 $ 0.30\pm 0.11$&
 $ 0.25\pm 0.43$&
 $ 0.89\pm 0.09$&
 $ 0.96\pm 0.10$&
 3.800\\
 060607A&
 2&
 $ 0.47\pm 0.05$&
 $ 0.94\pm 0.10$&
 $ 0.80\pm 0.06$&
 $ 0.86\pm 0.09$&
 3.082\\
 060614&
 1&
 $ 0.96\pm 0.03$&
 $ 0.88\pm 0.12$&
 $ 0.92\pm 0.08$&
 $ 0.77\pm 0.12$&
 0.125\\
 060707&
 1&
 $ 0.75\pm 0.10$&
 $ 0.84\pm 0.24$&
 $ 1.03\pm 0.13$&
 $ 1.20\pm 0.46$&
 3.430\\
 060708&
 1&
 $ 0.66\pm 0.08$&
 $ 2.22\pm 0.51$&
 $ 1.73\pm 0.20$&
 $ 1.47\pm 0.16$&
 \\
 060712&
 1&
 $ 0.82\pm 0.20$&
 $ 2.07\pm 0.33$&
 $ 1.36\pm 0.34$&
 $ 1.75\pm 0.23$&
 \\
 060714&
 1&
 $ 0.95\pm 0.07$&
 $ 1.36\pm 0.31$&
 $ 1.12\pm 0.12$&
 $ 1.36\pm 0.27$&
 2.710\\
 060717&
 1&
 $ 0.68\pm 0.26$&
 $ 1.23\pm 0.42$&
 &
 &
 \\
 060719&
 1&
 $ 1.03\pm 0.08$&
 $ 1.47\pm 0.62$&
 $ 1.36\pm 0.11$&
 $ 1.37\pm 0.15$&
 \\
 060729&
 1&
 $ 0.83\pm 0.10$&
 $ 1.64\pm 0.05$&
 $ 1.42\pm 0.04$&
 $ 1.39\pm 0.05$&
 0.540\\
 060801&
 3&
 $-0.55\pm 0.18$&
 $ 0.51\pm 0.32$&
 &
 &
 \\
 \enddata
 \end{deluxetable}

%% file: tab4.tex
 \begin{deluxetable}{lrrrccc}
 \tablewidth{0pt}
 \tablecaption{
 Potential jet breaks. Jet break times, $T_{j}$, in seconds.
 For the top 8 GRBs the break must have occurred at or before the
 end of the plateau and the jet break time given is $T_{a}$.
 For the bottom 4 the break was observed in
 the decay after the plateau and is listed in Table \ref{tab2}.
 $\alpha_{jd}$ and $\beta_{jd}$ are the indices measured after
 the break.
 The estimated jet angle $\theta_{j}$ (degrees) and
 jet energy $\log_{10}(E_{\gamma}\: ergs)$
 are listed for GRBs with measured redshifts, $z$.\label{tab4}}
 \tablehead{
 \colhead{GRB}&\colhead{$\log_{10}(T_{j})$}&
 \colhead{$\alpha_{jd}$}&\colhead{$\beta_{jd}$}&
 \colhead{$z$}&
 \colhead{$\theta_{j}$}&\colhead{$\log_{10}(E_{\gamma})$}}
 \startdata
 050603&
 $  4.83_{\;  3.69}^{\;  5.13}$&
 $  1.74_{\;  1.52}^{\;  2.21}$&
 $ 0.70\pm 0.10$&
 2.821&
   3.0&
  50.8\\
 050730&
 $  4.13_{\;  4.07}^{\;  4.21}$&
 $  2.74_{\;  2.57}^{\;  2.98}$&
 $ 0.62\pm 0.05$&
 3.970&
   1.6&
  49.9\\
 050802&
 $  3.96_{\;  3.86}^{\;  4.05}$&
 $  1.66_{\;  1.54}^{\;  1.79}$&
 $ 0.81\pm 0.09$&
 1.710&
   2.2&
  49.4\\
 060413&
 $  4.58_{\;  4.44}^{\;  4.70}$&
 $  1.96_{\;  1.69}^{\;  2.26}$&
 $ 0.53\pm 0.12$&
 \\
 060421&
 $  3.04_{\;  2.54}^{\;  3.43}$&
 $  1.25_{\;  1.03}^{\;  1.55}$&
 $ 0.33\pm 0.37$&
 \\
 060526&
 $  3.84_{\;  3.51}^{\;  4.31}$&
 $  1.06_{\;  0.89}^{\;  1.37}$&
 $ 0.13\pm 0.62$&
 3.210&
   1.5&
  49.4\\
 060605&
 $  4.16_{\;  4.03}^{\;  4.29}$&
 $  2.04_{\;  1.76}^{\;  2.39}$&
 $ 0.96\pm 0.10$&
 3.800&
   2.2&
  49.3\\
 060614&
 $  5.00_{\;  4.91}^{\;  5.08}$&
 $  1.98_{\;  1.78}^{\;  2.22}$&
 $ 0.77\pm 0.12$&
 0.125&
  10.8&
  49.5\\
 \hline
 050315&
 $  5.48_{\;  5.10}^{\;  5.75}$&
 $  2.00_{\;  0.86}^{\;  4.07}$&
 $ 1.29\pm 0.31$&
 1.949&
   6.8&
  50.9\\
 050814&
 $  4.93_{\;  4.41}^{\;  5.36}$&
 $  1.80_{\;  1.01}^{\;  3.16}$&
 $ 0.71\pm 0.10$&
 5.300&
   2.7&
  50.7\\
 060105&
 $  5.04_{\;  4.66}^{\;  5.26}$&
 $  3.56_{\;  1.45}^{\;  8.29}$&
 $ 1.07\pm 0.03$&
 \\
 060607A&
 $  4.81_{\;  4.06}^{\; 63.51}$&
 $  4.47_{\;  0.00}^{\;  8.94}$&
 $ 0.86\pm 0.09$&
 3.082&
   3.4&
  50.3\\
 \enddata
 \end{deluxetable}

%% file: tab5.tex
 \begin{deluxetable}{lccc}
 \tablewidth{0pt}
 \tablecaption{
 Comparison of the final X-ray decay index, $\alpha_{ad}$,
 with the optical decay index, $\alpha_{opt}$, for
 afterglows with no jet break.
 $^{\ast}$This error is dominated by systematic curvature
 of the decay rather than
 photometric accuracy.\label{tab5}}
 \tablehead{
 \colhead{GRB}&\colhead{$T_{a}$ secs}&
 \colhead{$\alpha_{ad}$ (90\%)}&\colhead{$\alpha_{opt}$}}
 \startdata
 050525A & $8.3\times10^2$ & 1.30-1.48 & $1.46\pm0.15$ \\
 050820A & $9.1\times10^3$ & 1.14-1.21 & $1.03\pm0.04$ \\
 060206 & $7.2\times10^3$ & 1.18-1.29 & $1.26\pm0.2^{\ast}$ \\
 060707 & $3.8\times10^3$ & 0.75-0.91 & $0.97\pm0.09$ \\
 060729 & $1.2\times10^5$ & 1.19-1.34 & $1.27\pm0.03$ \\
 \enddata
 \end{deluxetable}